\documentclass[journal]{IEEEtran}

\makeatletter
\def\ps@headings{%
	\def\@oddhead{\mbox{}\scriptsize\rightmark \hfil \thepage}%
	\def\@evenhead{\scriptsize\thepage \hfil \leftmark\mbox{}}%
	\def\@oddfoot{}%
	\def\@evenfoot{}}
\makeatother 

\usepackage{graphicx}
\usepackage{amsmath, amsfonts,epsfig, multirow, floatflt}
\usepackage{amssymb}
\usepackage{subfigure}
\usepackage{cite}
\usepackage{algorithm,algorithmic}
\usepackage{hyperref}
\usepackage{enumitem}
\usepackage{array}
\usepackage{color}
\usepackage[table]{xcolor}

\begin{document}
	
\title{Multi-user Resource Control with Deep Reinforcement Learning in IoT Edge Computing}

\author{Lei~Lei {\it Senior Member, IEEE}, Huijuan~Xu, Xiong~Xiong, Kan~Zheng {\it Senior Member, IEEE}, Wei~Xiang {\it Senior Member, IEEE}, and Xianbin~Wang {\it Fellow, IEEE}}

\maketitle

\begin{abstract}
	By leveraging the concept of mobile edge computing (MEC), massive amount of data generated by a large number of Internet of Things (IoT) devices could be offloaded to MEC server at the edge of wireless network for further computational intensive processing. However, due to the resource constraint of IoT devices and wireless network, both the communications and computation resources need to be allocated and scheduled efficiently for better system performance. In this paper, we propose a joint computation offloading and multi-user scheduling algorithm for IoT edge computing system to minimize the long-term average weighted sum of delay and power consumption under stochastic traffic arrival. We formulate the dynamic optimization problem as an infinite-horizon average-reward continuous-time Markov decision process (CTMDP) model. One critical challenge in solving this MDP problem for the multi-user resource control is the curse-of-dimensionality problem, where the state space of the MDP model and the computation complexity increase exponentially with the growing number of users or IoT devices. In order to overcome this challenge, we use the deep reinforcement learning (RL) techniques and propose a neural network architecture to approximate the value functions for the post-decision system states. The designed algorithm to solve the CTMDP problem supports semi-distributed auction-based implementation, where the IoT devices submit bids to the BS to make the resource control decisions centrally. Simulation results show that the proposed algorithm provides significant performance improvement over the baseline algorithms, and also outperforms the RL algorithms based on other neural network architectures. 
\end{abstract}

	
	\section{Introduction}
	Mobile edge computing (MEC) is an emerging technology that provides cloud computing capabilities at the edge of the mobile networks in close proximity to the mobile subscribers. Compared with mobile cloud computing (MCC), MEC can reduce latency and offer an improved user
	experience. On the other hand, the Internet of Things (IoT) comprises IoT devices with sensing, actuating, computation, and communication capabilities, which are connected into the Internet and collaboratively enable a wide of variety of new applications, including smart city/home, e-health, and industrial automation. As the IoT devices normally have very limited computation and storage capabilities, MEC enables the latency-sensitive IoT applications to offload the huge amount of sensed data to the MEC servers, which are deployed near the base stations (BSs) and offer large storage and computation facilities \cite{IoT:Pan,IoT:Abbas,Access:Yu,IoT:Premsankar}. To upload the sensed data from the IoT devices to the MEC server, NB-IoT cellular transmission technology is an attractive option, which is recently introduced in Third Generation Partnership Project (3GPP) Release 13, and is a long-term evolution (LTE) variant designed specifically for IoT \cite{IoT:Chen,IoT:Xu}. It enables mobile operators to efficiently support a massive number of IoT devices with low data rate transmissions and improved coverage using a small portion of their existing available licensed spectrum. NB-IoT has received great interest from major industrial partners in 3GPP, such as Ericsson, Nokia, Intel and Huawei \cite{Access:Malik}.\par
	
	In this paper, we consider an NB-IoT edge computing system, where MEC servers are deployed at NB-IoT enabled BSs. Based on this system, mobile operators can provide an efficient solution to the IoT applications by jointly optimize the radio and computational resources. One important challenge in the resource control for such a system is the offloading problem, which decides whether an IoT device should offload a chunk of sensed data to the MEC server or not. Offloading reduces the data computation delay as the central processing units (CPUs) of the MEC servers are much faster than those of the IoT devices, but it also incurs additional delay from data transmission. Moreover, the power consumption of local computation versus wireless transmission for an IoT device usually needs to be considered as well, as many IoT devices have limited energy (e.g., powered by batteries). On the other hand, the radio resource allocation decisions in NB-IoT will have significant effects on the data transmission delay and power consumption, which in turn affect the offloading performance.  \par
	
	\subsection{Related Work}
	The joint radio and computational resource control problem in multi-user MEC system has been studied in a few recent literatures, where several mobile devices share the same MEC server. A survey is provided in \cite{Survey:Mao}, where the computation task models considered in the existing research works are divided into deterministic versus stochastic. The deterministic task models consider that no new task will arrive until the old task is executed or discarded, so that the resource control decision of a particular task is made solely based on the information of the current task \cite{Access:Zhang,ICC:Chen}. On the other hand, the stochastic task models are more practical and consider that the tasks arrive according to a stochastic process and are buffered in a queue if cannot be processed immediately upon arrival. The resource control decisions for a particular task under the stochastic task models need to consider their impacts on the future tasks in terms of the long-term average performance of the system. Therefore, the problem is more complex under the stochastic task models, especially in the multi-user scenario due to the large dimensionality of the problem. A solution using the Lyapunov Optimization method is given in \cite{TWC:Mao} which considers a general wireless network and optimizes the energy consumption. In \cite{JSAC:Lyu}, a perturbed Lyapunov function is designed to stochastically maximize a network utility balancing throughput and fairness, and a knapsack problem is solved per slot for the optimal offloading schedule.\par 
	
	Markov Decision Process (MDP) is a powerful dynamic optimization theory to obtain the optimal resource control policy under the stochastic task arrival model in terms of the long-term average performance. However, solving the MDP model for the multi-user system is difficult due to the well-known curse-of-dimensionality problem, where the state space grows exponentially with the number of users \cite{Cui:TSP,Lei:TWC}. For this reason, previous studies based on the MDP models are mainly restricted to the single user MEC system \cite{ISIT:Liu,Globecom:Chen,Secon:Hong}. On the other hand, reinforcement learning (RL), especially deep reinforcement learning (DRL), provides a class of solution methods to address the curse-of-dimensionality problem in MDP, where the agents interact with the environment to learn optimal policies that map states to actions \cite{Sutton:book}. DRL algorithm can be broadly classified into value-based method, such as DQN \cite{Volodymyr:Nature}; policy gradient method; and actor-critic method which can be considered as a combination of value-based and policy gradient methods. The DRL algorithms enable RL to scale to problems that were previously intractable. Recent years have seen increasing applications of RL \cite{Lei:TWC,ISIT:Liu,Globecom:Chen,Secon:Hong,Lei:IoT} and DRL \cite{Chen:IoT,Wei:IoT,He:TET,Wei:ICC,Huang:DCN,Li:WCNC,Chen:arxiv,Wang:ETT} algorithms on the resource control problems in the MEC and IoT systems. \par
	
	Specifically, DRL algorithms for multi-user MEC system have been considered in several existing works. \cite{Huang:DCN} and \cite{Li:WCNC} focus on the offloading and resource allocation problems under deterministic task models, where a fixed number of tasks per user need to be processed either locally or offloaded to the edge server. DQN based techniques are applied to solve the respectively problems. This is different from the stochastic task model considered in this paper. In \cite{Chen:arxiv}, distributed power allocation policies for local execution and computation offloading are derived under stochastic task model with dynamic task arrival process by applying the deep deterministic policy gradient (DDPG) algorithm. It is considered in \cite{Chen:arxiv} that all the users can transmit simultaneously by leveraging multi-user MIMO. This is different from the consideration in this paper for NB-IoT system, where only one user can be scheduled for transmission over the $180$ kHz bandwidth. The mutual exclusion nature in multi-user resource allocation makes it hard to design a fully distributed solution as in \cite{Chen:arxiv}, where each user makes independent decisions according to its local state information. Moreover, the offloading and resource allocation problem in \cite{Chen:arxiv} is reduced to power allocation problem by considering a data-partition task model \cite{Survey:Mao}, which results in a continuous action space that favors policy gradient or actor-critic algorithms over value-based algorithms. In this paper, we adopt the value-based algorithm as the action space is discrete. \par   
	
	Another thread of related research is the multi-agent RL \cite{Littman:ML}, which typically involves multiple agents learning individual policies. The state transitions and rewards depend on the joint actions of all the agents. Compared with single-agent RL, multi-agent RL can solve the action space explosion problem, i.e., the cardinality of action space grows exponentially with the number of agents. For example, independent-Q learning is a popular algorithm in which each agent independently learns its own policy, treating other agents as part of the environment \cite{Tan:ML}. However, a problem with independent-Q learning is that the environment becomes non-stationary \cite{Foerster:arxiv}. There are several survey papers on multi-agent RL that introduce the challenges and solutions \cite{Nguyen:arXiv,Busoniu:TSMC,Hernandez-Leal:arxiv}. In this paper, multi-agent RL algorithms cannot be applied directly because of the mutual exclusion nature of the resource allocation problem. As the radio resources can only be allocated to at most one user at a time, each agent cannot make individual decisions ignoring the decisions of the other agents. Moreover, due to the semi-Markov characteristics of the RL model, the action space does not grow exponentially with the number of users as in multi-agent RL. At each decision epoch, only the offloading decision of one user needs to be considered upon the arrival of a new task. \par

\subsection{Contributions}
	In this paper, we propose a deep reinforcement learning method with the value function approximation architecture based on ANNs for the multi-user resource control problem of the NB-IoT edge computing system. We formulate the dynamic optimization problem as an infinite-horizon average-reward continuous-time Markov decision process (CTMDP) model. In the CTMDP model, the global reward function can be represented as the sum of local reward functions per user. This corresponds to a typical optimization objective for multi-user resource control problem, where the overall system performance, e.g., delay, power consumption, is the sum or average value of the per-user performance. Moreover, the resource control action includes the offloading action and multi-user scheduling action. The latter has the constraint that at most one user can be scheduled for data transmission at a time. This is a typical intra-cell resource allocation consideration in cellular networks, which makes it difficult to directly apply existing multi-agent RL algorithms. \par

    \emph{The main contribution of this paper lies in the design of a neural network architecture for function approximation that facilitates semi-distributed implementation of the learning algorithm in the multi-user environment. Specifically, the edge server and BS make the resource control decisions with an auction-based mechanism, where the large amount of IoT devices distributively compute and submit bids to the BS and edge server. }\par
    
    The motivation for semi-distributed implementation is twofold. Firstly, although the proposed algorithm can be implemented centrally at the BS, the computation complexity and required storage capacity increase with the increasing number of IoT devices. Therefore, by efficient collaboration between BS and IoT devices, the IoT devices can help to alleviate the computational and storage burdens from the BS. This is in accordance with the design principles for new generation of wireless networks - making use of smart user equipments (UEs) to help the BS. Secondly, although a fully distributed implementation seems attractive from performance perspective, the mobile operators need to be able to control the scarce spectrum resources in the license band \cite{wirelessMag:Lei}. Therefore, in the proposed semi-distributed implementation, the BS makes control decisions while the IoT devices submit individual bids. \par  
    
   In the design of neural network architecture, we propose several novel features to facilitate semi-distributed implementation with good performance and limited communications overhead. Firstly, we approximate the global value function by the summation over all the users of their respective product of local value function and local feature. The local value function depends solely on the local system state of a user. On the other hand, the local feature depends on the global system state to improve the accuracy of approximation. Secondly, we adopt a convolutional layer to compress the local system state of every user to a single scalar. This can greatly reduce the signaling overhead for the BS to inform IoT devices of the global system state as well as improving the performance of the learning algorithm. Thirdly, we insert a multiplication layer before the output layer so that only the local value function associated with the current local system state needs to be updated per decision epoch for each user. This greatly reduces the computation complexity and signaling overhead associated with parameter update. Finally, with the auction-based mechanism in implementation, each IoT device submits a bid per local action, and the BS selects the joint action that results in the optimum global value function. In this way, global optimum is ensured through semi-distributed implementation. The proposed function approximation architecture can be adopted by other multi-user resource control problems that share similar problem structure. \par

	The rest of the paper is organized as follows. In Section II, the system model is introduced. Section III formulates the CTMDP problem, which is solved in Section IV using the value function approximation, neural networks, and reinforcement learning techniques. The semi-distributed implementation procedure is also discussed in Section IV. In Section V, the performances of the proposed algorithm are compared with those of the baseline algorithms as well as the other DRL algorithms by simulation. Section VI concludes the paper. \par

	\section{System Model}
\begin{figure}[!htb]
		\centering
		\includegraphics[angle=90,width=0.5\textwidth]{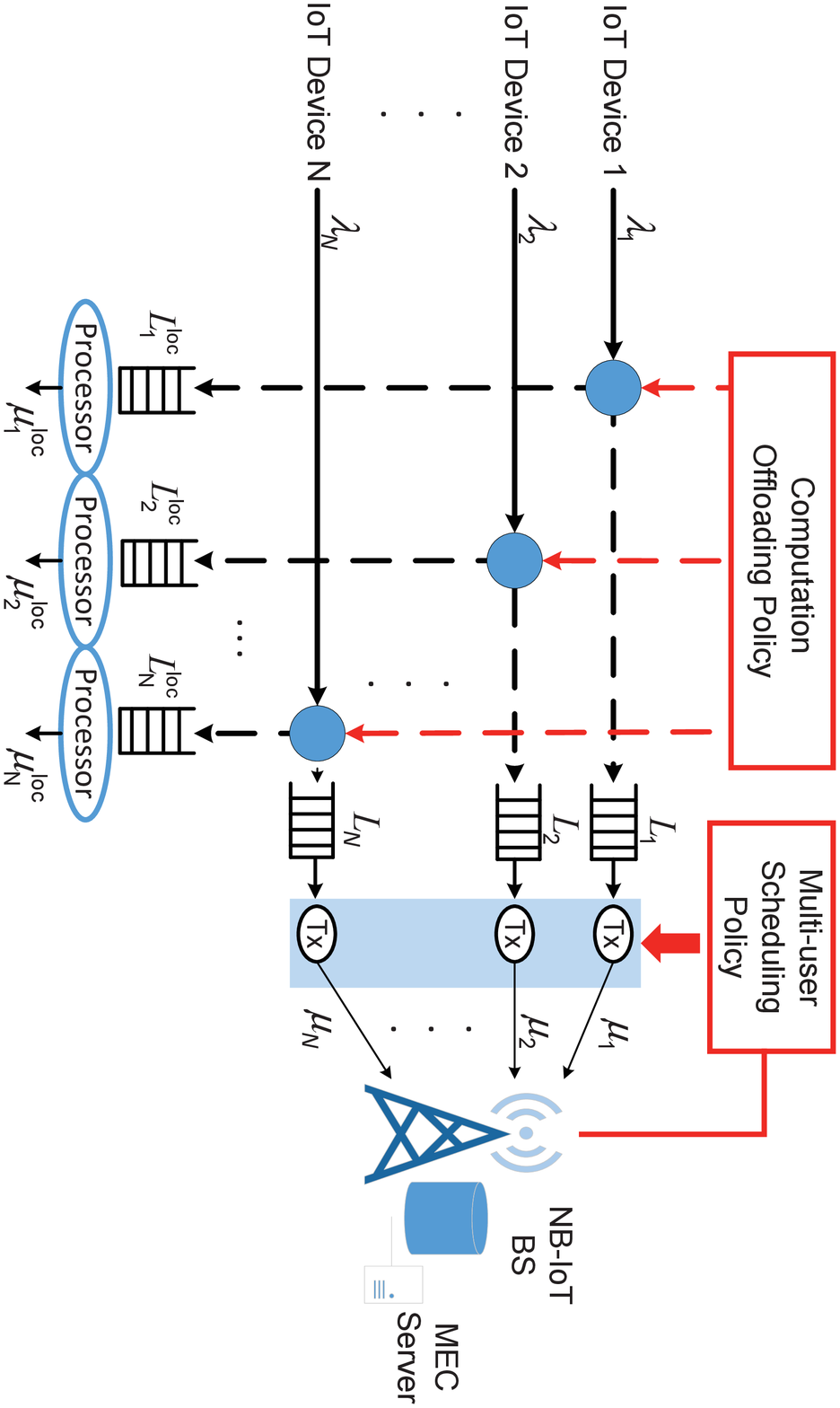}
		\caption{IoT edge computing system model.}
		\label{queue_mod}
	\end{figure}

	We consider an IoT edge computing system, where a BS with an MEC server serves $N$ IoT devices in a singel cell \cite{Lei:IoT}. For each IoT device $n\in\mathcal{N}=\{1,2,\cdots,N\}$, the sensed data arrives in packets according to a Poisson distribution of mean arrival rate $\lambda_{n}$. There are two queues for each IoT device to buffer the sensed packets. One is the transmission queue for the packets that are to be offloaded to the MEC server for remote computation, and the other is the processing queue for the packets that are be locally processed by the IoT device. When a new packet arrives at an IoT device, the offloading function decides whether to place it in the transmission queue for offloading, or in the processing queue for local processing. Moreover, the multi-user scheduling function in the wireless network decides how to allocate the radio resources to different IoT devices for the transmission of the offloaded packets. The system model for the IoT edge computing system considered in this paper is illustrated in Fig.\ref{queue_mod}.\par
	
   	\newtheorem{assumption}{Assumption}
   	\begin{assumption}[Resource Unit (RU) configuration in NB-IoT]    
     We consider that the RU configuration with $12$ subcarriers $\times$ $2$ time slots is always selected for every IoT device \cite{VTC:Ratasuk}. Therefore, only one IoT device can be scheduled for transmission at the same time.
   	\end{assumption}

   \newtheorem{assumption2}[assumption]{Assumption}
   \begin{assumption2}[Link adaption in NB-IoT]    
In LTE system, link adaptation is performed dynamically per $1$ms subframe to adapt the Modulation and Coding Scheme (MCS) level according to the instantaneous channel quality. As a narrowband transmission technology with a relatively low data rate, the transmission of a transport block (TB) in NB-IoT can occupy multiple consecutive subframes, i.e., a TB may be mapped to $\{1,2,3,4,5,6,8,10\}$ RUs in time \cite{VTC:Ratasuk}. This means that the transmission duration can be larger than the coherence time of the wireless channel. Therefore, in this paper, we consider that the link adaptation is performed according to the time-average wireless channel conditions of the IoT devices determined only by the large-scale fading effects, i.e., pathloss and shadowing.  Moreover, we focus on those IoT applications where the locations of the IoT devices will not often change once they are deployed, e.g., smart metering. Therefore, the MCS level and the corresponding transmission data rate for an IoT device will remain the same as long as it does not change its location.
   \end{assumption2} 
  
 	
As a narrowband transmission technology with a relatively low data rate, the transmission of a transport block in NB-IoT can occupy multiple consecutive subframes \cite{VTC:Ratasuk}. In this paper, we consider that the transmission duration of a packet is exponentially distributed with a mean value $1/\mu_{n}$, where $\mu_{n}$ is the mean transmission rate in terms of packets per second for IoT device $n\in\mathcal{N}$. Moreover, the power consumption is a constant value $P_{n}$ for any IoT device $n\in\mathcal{N}$ \cite{Access:Yu2}.\par

We consider that the mean local processing time of IoT device $n$ is exponentially distributed with a mean of $1/\mu_{n}^{\mathrm{loc}}$, where $\mu_{n}^{\mathrm{loc}}$ is the mean processing rate in terms of packets per second for IoT device $n\in\mathcal{N}$. The power consumption for processing the sensed data locally at the IoT device is a constant value denoted by $P_{n}^{\mathrm{loc}}$ \cite{Globecom:Chen}.\par

%
	
	In this paper, we will try to jointly derive the optimal scheduling policy and computation offloading policy that minimizes the weighted sum of the average delay and power consumption over all the IoT devices. Specifically, the average delay depends on the delay values for both the offloaded as well as the un-offloaded packets. The delay of an offloaded packet includes three parts, i.e., the uplink transmission delay, the remote computation delay, and possibly the downlink transmission delay. We make the following assumption when deriving the average delay for the offloaded packets.
	\newtheorem{assumption3}[assumption]{Assumption}
	\begin{assumption3}[Delay for offloaded packets]    
	We assume that the average delay for the offloaded packets equal to their average uplink transmission delay. This is because the sum of the remote computation delay and the downlink transmission delay is usually neligible compared with the uplink transmission delay and local computation delay due to much more powerful CPUs of the MEC servers and much heavier uplink IoT traffic.
	\end{assumption3}

	\section{CTMDP Model}
	In this section, we shall formulate an infinite horizon average reward Continuous Time Markov Decision Process (CTMDP) problem to minimize the weighted sum of the average delay and power consumption for the IoT devices.\par

	\subsection{Global System State}
	We formulate a CTMDP model where the global system states are observed at each packet arrival and departure event. We denote the global system state at the $k$-th decision epoch, $\forall k\in\{0,1,2,\cdots\}$, by $\mathbf{s}_{k}=(\mathbf{L}_{k},\mathbf{L}_{k}^{\mathrm{loc}},e_{k},b_{k})$. 
	
	\paragraph{Transmission queue state $\mathbf{L}_{k}$}
	$\mathbf{L}_{k}=\{L_{n,k}\}_{n=1}^{N}$ is the vector of transmission queue length observed at the beginning of the $k$-th decision epoch when the packet arrival/departure event has just occurred. $L_{n,k}\in\{0,1,\cdots,M\}$, $\forall n\in\mathcal{N}$ denotes the transmission queue length of IoT device $n$, where $M$ is the maximum transmission queue length. 
	\paragraph{Processing queue state $\mathbf{L}_{k}^{\mathrm{loc}}$}	
	$\mathbf{L}_{k}^{\mathrm{loc}}=\{L_{n,k}^{\mathrm{loc}}\}_{n=1}^{N}$ is the vector of processing queue length observed at the beginning of the $k$-th decision epoch, where $L_{n,k}^{\mathrm{loc}}\in\{0,1,\cdots,M^{\mathrm{loc}}\}$, $\forall n\in\mathcal{N}$ denotes the processing queue length of IoT device $n$. $M^{\mathrm{loc}}$ is the maximum processing queue length.
	\paragraph{Event $e_{k}$}
	Let $e_{k}\in-\mathcal{N}\cup\{0\}\cup\mathcal{N}$ indicate the event occurred at the beginning of the $k$-th decision epoch which triggers the state transition from $\mathbf{s}_{k-1}$ to $\mathbf{s}_{k}$.
	\begin{description}[leftmargin=5em,style=nextline]
		\item[$e_{k}\in \mathcal{N}$] represents a packet arrival at IoT device $e_{k}$;
		\item[$e_{k}=0$] represents a packet departure from the scheduled transmission queue;
		\item[$e_{k}\in-\mathcal{N}$] represents a packet departure from the processing queue of IoT device $-e_{k}$, where $-\mathcal{N}=\{-1,-2,\cdots,-N\}$.
	\end{description}
	\paragraph{Scheduled transmission queue $b_{k}$}
	$b_{k}\in\{0\}\cup\mathcal{N}$ is the scheduling action at the last (i.e., $(k-1)$-th) decision epoch.
	\begin{description}[leftmargin=5em,style=nextline]
	\item[$b_{k}\in\mathcal{N}$] represents the index of the scheduled transmission queue at the last (i.e., $(k-1)$-th) decision epoch;	
	\item[$b_{k}=0$] means no transmission queue is scheduled.
	\end{description}
	
	\newtheorem{example}{Example}
	\begin{example}[Definition of global system state]
	The global system state $(\mathbf{0},\mathbf{0},3,0)$ ($\mathbf{0}$ denotes a $1\times N$ vector of zeros) indicates that the system state transits to the current state due to a packet arrival at IoT device $3$. At the beginning of the current system state, all the transmission queues and processing queues are empty. No transmission queue is scheduled at the previous system state. 
	\end{example}

    The cardinality of the global system state space $\mathcal{S}$ is $|\mathcal{S}|=(M+1)^{N}\times(M^{\mathrm{loc}}+1)^{N}\times(2N+1)\times(N+1)$, which grows exponentially with the number of IoT devices $N$. \par
	
	\subsection{Action}
	When a system state transition occurs due to a packet arrival or departure event, an action will be taken in the CTMDP model. Define the action at the $k$-th decision epoch as $\mathbf{a}_{k}=(a_{\mathrm{o},k},a_{\mathrm{s},k})$.\par 
	
	\paragraph{Offloading action $a_{\mathrm{o},k}$}
	$a_{\mathrm{o},k}\in\mathcal{A}_{\mathrm{o}}=\{-1,0,1\}$ represents the offloading action, which is only performed when there is a packet arrival.
	\begin{description}[leftmargin=5em,style=nextline]
	\item[$a_{\mathrm{o},k}=1$]	means the newly arrived packet is offloaded;
	\item[$a_{\mathrm{o},k}=0$] means the newly arrived packet is not offloaded; or an offloading action is not applicable in the current system state (i.e., when $e_{k}$ in the current system state is a packet departure event);
	\item[$a_{\mathrm{o},k}=-1$] means the newly arrived packet is dropped because both the transmission queue and processing queue of the IoT device are saturated.
	\end{description}
    From the above definition, it is obvious that the offloading action space is dependent on the system state. This dependency is further demonstrated by the fact that if one of two queues (i.e., transmission queue and processing queue) of the IoT device at which the packet arrived is saturated, the packet can only be dispatched to the other queue, and thus the offloading action is determined. Therefore, the state-dependent offloading action space is given as
	\begin{equation}
	\label{eq4}
	\mathcal{A}_{\mathrm{o},\mathbf{s}_{k}}=\left\{
	\begin{array}{ll}
	\{0\}, & e_{k}\in-\mathcal{N}\cup\{0\}, \mathrm{or}, \\ 
	& e_{k}=n\in\mathcal{N}, L_{n,k}=M, L_{n,k}^{\mathrm{loc}}<M^{\mathrm{loc}}   \\
	\{1\}, & e_{k}=n\in\mathcal{N}, L_{n,k}<M, L_{n,k}^{\mathrm{loc}}=M^{\mathrm{loc}}   \\
	\{-1\}, & e_{k}=n\in\mathcal{N}, L_{n,k}=M, L_{n,k}^{\mathrm{loc}}=M^{\mathrm{loc}}   \\
	\{0,1\},
	& \mathrm{otherwise} \\	
	\end{array}\right. .
	\end{equation}
	
	After the offloading action is made in the $k$-th decision epoch, the arrived packet will be dropped or added to the transmission queue or processing queue of IoT device $e_{k}$ depending on the offloading action. Therefore, the processing and transmission queue length of IoT device $e_{k}$ can be different from the values of $L_{e_{k},k}$ and $L_{e_{k},k}^{\mathrm{loc}}$ in the system state. Let $\tilde{L}_{n,k}^{\mathrm{loc}}$ and $\tilde{L}_{n,k}$ denote the processing and transmission queue length of any IoT device $n\in\mathcal{N}$ during the $k$-th decision epoch after the offloading decision is made with system state $\mathbf{s}_{k}$, we have 
	\begin{equation}
	\label{eq6}	
	\tilde{L}_{n,k}^{\mathrm{loc}}=\left\{
	\begin{array}{ll}
	L_{n,k}^{\mathrm{loc}}+1, & e_{k}=n\in\mathcal{N}, a_{\mathrm{o},k}=0 \\ 
	L_{n,k}^{\mathrm{loc}} & \mathrm{otherwise}	
	\end{array}\right. ,
	\end{equation}	
	
	\noindent and
	\begin{equation}
	\label{eq7}	
	\tilde{L}_{n,k}=\left\{
	\begin{array}{ll}
	L_{n,k}+1, & e_{k}=n\in\mathcal{N}, a_{\mathrm{o},k}=1 \\ 
	L_{n,k} & \mathrm{otherwise}	
	\end{array}\right. .
	\end{equation}
	
	Define the post-decision transmission and processing queue vectors at the $k$-th decision epoch as $\tilde{\mathbf{L}}_{k}=\{\tilde{L}_{n,k}\}_{n=1}^{N}$ and $\tilde{\mathbf{L}}_{k}^{\mathrm{loc}}=\{\tilde{L}_{n,k}^{\mathrm{loc}}\}_{n=1}^{N}$, based on the values of which the scheduling action $a_{\mathrm{s},k}$ will be made for the $k$-th decision epoch. \par
	
	\paragraph{Scheduling action $a_{\mathrm{s},k}$}
	$a_{\mathrm{s},k}\in\mathcal{A}_{\mathrm{s}}= \{0\}\cup\mathcal{N}$ is the scheduling action.
	\begin{description}[leftmargin=5em,style=nextline]
		\item[$a_{\mathrm{s},k}\in\mathcal{N}$] represents the index of the transmission queue that is scheduled at the current ($k$-th) decision epoch;
		\item[$a_{\mathrm{s},k}=0$] means that no queue is scheduled, which only happens when all the transmission queues are empty at the time the scheduling action is determined, i.e., $\tilde{\mathbf{L}}_{k}=\mathbf{0}$.;
	\end{description}
	
	In this paper, we consider non-preemptive scheduling. Therefore, the scheduling action is only updated when (1) there is a packet departure from a transmission queue (i,e, $e_{k}=0$); (2) no queue is scheduled at the time that an arrival event occurs (i.e., $b_{k}=0$ and $e_{k}\in\mathcal{N}$). In either case, the scheduled IoT device is selected from the the set of IoT devices with non-empty transmission queues, i.e., $\mathcal{N}_{\mathrm{s},k}=\{n|\tilde{Q}_{n}\neq 0,n\in\mathcal{N}\}$.  Otherwise, the scheduling action remains the same as the previous decision epoch (i.e., $a_{\mathrm{s},k}=b_{k}$).
	Therefore, the state-dependent scheduling action set is given as
	\begin{equation}
	\label{eq5}
	\mathcal{A}_{\mathrm{s},\mathbf{s}_{k}}=\left\{
	\begin{array}{ll}
	\{0\} & \tilde{\mathbf{L}}_{k}=\mathbf{0} \\
	\mathcal{N}_{\mathrm{s},k} & \tilde{\mathbf{L}}_{k}\neq\mathbf{0}, \mathrm{and} \\
	                       &e_{k}=0 \ \mathrm{or} \ (e_{k}\in\mathcal{N},b_{k}=0) \\
	\{b_{k}\}, & \mathrm{otherwise} \\
	\end{array}\right. .
	\end{equation}
	
 Note that the cardinalities of the offloading action space and scheduling action space are $3$ and $N+1$, respectively.
	
	\subsection{Post-Decision Global System State}
    We define the post-decision global system state at the $k$-th decision epoch as $\tilde{\mathbf{s}}_{k}$, which is a deterministic function of the global system state $\mathbf{s}_{k}$ and the action $\mathbf{a}_{k}$ at the $k$-th decision epoch as below:
	\begin{equation}
	\label{eq8}
	\tilde{\mathbf{s}}_{k}=f(\mathbf{s}_{k},\mathbf{a}_{k})=(\tilde{\mathbf{L}}_{k},\tilde{\mathbf{L}}_{k}^{\mathrm{loc}},e_{k},a_{\mathrm{s},k}).
	\end{equation}
	\noindent Note that the state space of post-decision global system states is the same with that of the global system states as denoted by $\mathcal{S}$.\par
	
%
	\subsection{Transition Probability}
Given the global systems state $\mathbf{s}_{k}$ and action $\mathbf{a}_{k}$ at the $k$-th decision epoch, the transition to the global systems state $\mathbf{s}_{k+1}$ at the $(k+1)$-th decision epoch can be described in two phases.
	\begin{description}[leftmargin=10em,style=nextline]
		\item[Phase 1 $(\mathbf{s}_{k},\mathbf{a}_{k})\rightarrow\tilde{\mathbf{s}}_{k}$] where $\tilde{\mathbf{s}}_{k}$ can be derived by \eqref{eq8} as a deterministic function of $\mathbf{s}_{k}$ and $\mathbf{a}_{k}$ ;
		\item[Phase 2 $\tilde{\mathbf{s}}_{k}\rightarrow\mathbf{s}_{k+1}$] where $\mathbf{s}_{k+1}$ is a deterministic function of  $\tilde{\mathbf{s}}_{k}$ and $e_{k+1}$ as below:
    \end{description}	
		\begin{align}
\label{eq40}
\mathbf{s}_{k+1}&=h(\tilde{\mathbf{s}}_{k},e_{k+1}) \IEEEnonumber \\ 
&=\left\{
\begin{array}{ll}
(\tilde{\mathbf{L}}_{k},\tilde{\mathbf{L}}_{k}^{\mathrm{loc}},n,a_{\mathrm{s},k}), 
& e_{k+1}=n  \\
(\tilde{\mathbf{L}}_{\underline{a_{\mathrm{s},k}},k},\tilde{\mathbf{L}}_{k}^{\mathrm{loc}},0,a_{\mathrm{s},k}), 
& e_{k+1}=0 \\
(\tilde{\mathbf{L}}_{k},\tilde{\mathbf{L}}_{\underline{n},k}^{\mathrm{loc}},-n,a_{\mathrm{s},k}), 
& e_{k+1}=-n
\end{array}\right. n\in\mathcal{N} ,
\end{align}	
\noindent where $\tilde{\mathbf{L}}_{\underline{a_{\mathrm{s},k}},k}=\{\tilde{L}_{1,k},\cdots,\tilde{L}_{a_{\mathrm{s},k},k}-1,\cdots,\tilde{L}_{N,k}\}$, $\tilde{\mathbf{L}}_{\underline{n},k}^{\mathrm{loc}}=\{\tilde{L}_{1,k}^{\mathrm{loc}},\cdots,\tilde{L}_{n,k}^{\mathrm{loc}}-1,\cdots,L_{N,k}^{\mathrm{loc}}\}$. \par	
	
	Note that the event $e_{k+1}$ at the $(k+1)$-th decision epoch occurs when there is a packet arrival at any of the IoT devices, or there is a packet departure from the scheduled transmission queue, or from any of the non-empty processing queues. Therefore, the transition probabilities $q(\mathbf{s}_{k+1}|\tilde{\mathbf{s}}_{k})$ corresponds to the probabilities that event $e_{k+1}$ happens:	
	\begin{equation}
	\label{eq10}
	q(\mathbf{s}_{k+1}|\tilde{\mathbf{s}}_{k})=\left\{
	\begin{array}{ll}
	\frac{\lambda_{n}}{\tilde{\beta}(\tilde{\mathbf{s}}_{k})}, 
	&  \mathbf{s}_{k+1}=h(\tilde{\mathbf{s}}_{k},n), \\
	\frac{\mu_{a_{\mathrm{s},k}}}{\tilde{\beta}(\tilde{\mathbf{s}}_{k})}, 
	& \mathbf{s}_{k+1}=h(\tilde{\mathbf{s}}_{k},0), \\
	\frac{\mu_{n}^{\mathrm{loc}}\mathbf{1}_{\tilde{L}_{n,k}^{\mathrm{loc}}\neq 0}}{\tilde{\beta}(\tilde{\mathbf{s}}_{k})}, 
	& \mathbf{s}_{k+1}=h(\tilde{\mathbf{s}}_{k},-n),  \\
	0, 
	& \mathrm{otherwise}	
	\end{array}\right. ,
	\end{equation}	
	\noindent where we set $\mu_{0}=0$ as $e_{k+1}=0$ will not happen when no transmission queue is scheduled during the $k$-th decision epoch, i.e., $a_{\mathrm{s},k}=0$.\par
	
	The duration of the $k$-th decision epoch or equivalently, the sojourn time of the CTMDP in state $\mathbf{s}_{k}$ given action $\mathbf{a}_{k}$ is exponentially distributed with parameter $\beta(\mathbf{s}_{k},\mathbf{a}_{k})$ as\par
    \begin{equation}
	\label{eq9}
	\beta(\mathbf{s}_{k},\mathbf{a}_{k})=\sum_{n=1}^{N}\lambda_{n}+\mu_{a_{\mathrm{s},k}}+\sum_{n=1}^{N}\mu_{n}^{\mathrm{loc}}\mathbf{1}_{(\tilde{L}_{n,k}^{\mathrm{loc}}\neq 0)}, 
	\end{equation}
	\noindent where $\beta(\mathbf{s}_{k},\mathbf{a}_{k})=\tilde{\beta}(f(\mathbf{s}_{k},\mathbf{a}_{k}))=\tilde{\beta}(\tilde{\mathbf{s}}_{k})$ can also be expressed as a function of the post-decision state $\tilde{\mathbf{s}}_{k}$. \par

	\subsection{Reward Function}
	In order to derive the reward function of the CTMDP model, we first examine the optimization objective, which is to find the policy $\pi$ that minimizes the weighted sum of the average delay and power consumption over all the IoT devices. Note that a policy $\pi$ in an MDP model is a function that specifies the action $\mathbf{a}_{k}=\pi(\mathbf{s}_{k})$  that the decision maker will choose when in state $\mathbf{s}_{k}$. We formulate the above dynamic optimization problem as an average reward CTMDP problem.
	\newtheorem{problem}{Problem}
	\begin{problem}[average reward CTMDP problem to minimize the weighted sum of average delay and power consumption]
	\begin{align}
	\label{eq12}
	\min_{\pi}J(\pi)& =\sum_{n=1}^{N}\omega_{n}\bar{D}_{n}+\sum_{n=1}^{N}\gamma_{n}\bar{P}_{n}  \\ \nonumber 
	& =\lim_{T\rightarrow\infty} \frac{\mathbf{E}^{\pi}[\sum_{k=0}^{T}\int_{\sigma_{k}}^{\sigma_{k+1}}c(\mathbf{s}_{k},\pi(\mathbf{s}_{k}))dt]}{\mathbf{E}^{\pi}[\sum_{k=0}^{T}\tau_{k}]},
	\end{align}
	\noindent where $\omega_{n}$ and $\gamma_{n}$ on the RHS of the first equality are the weights for the average delay $\bar{D}_{n}$ and power consumption $\bar{P}_{n}$ of IoT device $n$, respectively. The weights $\omega_{n}$ and $\gamma_{n}$ indicate the relative importance of the average delay and power consumption of IoT device $n$ in the optimization problem. The RHS of the second equality is the classical form of an average reward CTMDP problem, where $\sigma_{k}$ and $\tau_{k}$ are the starting time and duration of the the $k$-th decision epoch. $c(\mathbf{s}_{k},\pi(\mathbf{s}_{k}))$ represents that a reward is incurred at this rate when the system is in state $\mathbf{s}_{k}$ and action $\pi(\mathbf{s}_{k})$ is chosen at the the $k$-th decision epoch. 
	\end{problem}

    From \eqref{eq12} and using Little's Law to derive the delay, we can derive the expression of $c(\mathbf{s}_{k},\pi(\mathbf{s}_{k}))$ as below:	
	\begin{align}
	\label{eq19}
	c(\mathbf{s}_{k},\pi(\mathbf{s}_{k}))=&\sum_{n=1}^{N}\big(\frac{\omega_{n}}{\lambda_{n}}(\tilde{L}_{n,k}+\tilde{L}_{n,k}^{\mathrm{loc}})+\gamma_{n} \IEEEnonumber \\ 
	&(P_{n}\times\mathbf{1}_{a_{\mathrm{s},k}=n}+P_{n}^{\mathrm{loc}}\times\mathbf{1}_{\tilde{L}_{n,k}^{\mathrm{loc}}\neq 0})\big),
	\end{align}

	\noindent where $\mathbf{1}_{\mathrm{condition}}$ is a random variable that takes the value of $1$ if the condition in the subscript is true, and $0$ otherwise. The detailed derivation is given in Appendix A.\par
	
	According to the CTMDP theory \cite{book:Puterman}, the reward function of the CTMDP model can be derived as
	\begin{equation}
	\label{eq20}	
	g(\mathbf{s}_{k},\pi(\mathbf{s}_{k}))=c(\mathbf{s}_{k},\pi(\mathbf{s}_{k}))/\beta(\mathbf{s}_{k},\pi(\mathbf{s}_{k})).
	\end{equation}
	\noindent where $g(\mathbf{s}_{k},\mathbf{a}_{k})=\tilde{g}(f(\mathbf{s}_{k},\mathbf{a}_{k}))=\tilde{g}(\tilde{\mathbf{s}}_{k})$ can also be expressed as a function of the post-decision state $\tilde{\mathbf{s}}_{k}$.\par	
	
	The optimal policy of the above CTMDP problem can be derived by solving the post-decision Bellman equation as
	\begin{align}
	\label{eq21}
	&V(\tilde{\mathbf{s}}_{k})=\sum_{\mathbf{s}_{k+1}\in\mathcal{S}}q(\mathbf{s}_{k+1}|\tilde{\mathbf{s}}_{k})\min_{\pi}\Big(\tilde{g}(\tilde{\mathbf{s}}_{k+1})+V\big(\tilde{\mathbf{s}}_{k+1}\big) \IEEEnonumber \\ 
	&-\theta/\tilde{\beta}(\tilde{\mathbf{s}}_{k+1})\Big),
	\end{align} 
	\noindent where $V(\cdot)$ is the post-decision global value function, and $\theta$ is the optimal average reward rate.\par

	\section{Solution by DRL}
	\subsection{Local System State}
	Define the local system state for IoT device $n$ at the $k$-th decision epoch as $s_{n,k}=(L_{n,k},L_{n,k}^{\mathrm{loc}},e_{n,k},b_{n,k})$. 
	\begin{description}
	\item[Local event $e_{n,k}$] $\in\{\mathrm{null},-1,0,1\}$, where
	\begin{description}[leftmargin=6em,style=nextline]
		\item[$e_{n,k}=1$] indicates a packet arrives at IoT device $n$;
		\item[$e_{n,k}=0$] indicates a packet departs from the transmission queue of IoT device $n$;
		\item[$e_{n,k}=-1$] indicates a packet departs from the processing queue of IoT device $n$;
		\item[$e_{n,k}=\mathrm{null}$] indicates the event at the $k$-th decision epoch does not happen at IoT device $n$.
    \end{description}	
	\item[Local schedule $b_{n,k}$] $\in\{0,1\}$, where
		\begin{description}[leftmargin=6em,style=nextline]
		\item[$b_{n,k}=1$] indicates that IoT device $n$ is scheduled at the $(k-1)$-th decision epoch;
		\item[$b_{n,k}=0$] indicates that IoT device $n$ is not scheduled at the $(k-1)$-th decision epoch.
		\end{description}	
    \end{description}		

Given the global system state $\mathbf{s}_{k}=(\mathbf{L}_{k},\mathbf{L}_{k}^{\mathrm{loc}},e_{k},b_{k})$ at the $k$-th decision epoch, the local system state at IoT device $n$ can be derived by 	
	\begin{equation}
	\label{eq22}
	e_{n,k}=\varepsilon_{n}(e_{k})=\left\{
	\begin{array}{ll}
	1 & e_{k}=n\in\mathcal{N},   \\
	0 & e_{k}=0, b_{k}=n, \\
	-1& e_{k}=-n\in-\mathcal{N},  \\
	\mathrm{null}& \mathrm{otherwise}
	\end{array}\right.,
	\end{equation}
	\begin{equation}
	\label{eq23}
	b_{n,k}=\mathbf{1}_{b_{k}=n}.
	\end{equation}
	The global system state $\mathbf{s}_{k}$ corresponds to the aggregation of the local system states $\{s_{n,k}\}_{n=1}^{N}$.

	\newtheorem{example2}[example]{Example}
    \begin{example2}[Definition of local system state]	
	Consider there are $3$ IoT devices in the system, and the global system state is $(\{2,3,1\},\{2,0,2\},3,2)$. Thus, the local system states at the $3$ IoT devices are $(2,2,\mathrm{null},0)$, $(3,0,\mathrm{null},1)$, and $(1,2,1,0)$, respectively. 
	\end{example2}

	Given the local system state $s_{n,k}$ of IoT device $n$ at the $k$-th decision epoch, and the action $\mathbf{a}_{k}$, we define the post-decision local system state $\tilde{s}_{n,k}$, which can be derived by a deterministic function $f_{n}$ as
	\begin{equation}
	\label{eq24}
	\tilde{s}_{n,k}=f_{n}(s_{n,k},\mathbf{a}_{k})=(\tilde{L}_{n,k},\tilde{L}_{n,k}^{\mathrm{loc}},e_{n,k},\mathbf{1}_{a_{\mathrm{s},k}=n}).
	\end{equation}
	
	Given the post-decision local system state $\tilde{s}_{n,k}$ of IoT device $n$ at the $k$-th decision epoch, and the event $e_{k+1}$ at the $(k+1)$-th decision epoch, the local system state $s_{n,k+1}$ of IoT device $n$ at the $(k+1)$-th decision epoch can be derived by a deterministic function $h_{n}$ as below by combining \eqref{eq40} with the definitions of (post-decision) local system states:
	\begin{align}
	\label{eq41}
	&s_{n,k+1}=h_{n}(\tilde{s}_{n,k},e_{k+1})= \IEEEnonumber \\
	&\left\{
	\begin{array}{ll}
	(\tilde{L}_{n,k},\tilde{L}_{n,k}^{\mathrm{loc}},\varepsilon_{n}(e_{k+1}),\mathbf{1}_{a_{\mathrm{s},k}=n}) &  \varepsilon_{n}(e_{k+1}) \\
	& \in\{1,\mathrm{null}\},   \\
	(\tilde{L}_{n,k}-1,\tilde{L}_{n,k}^{\mathrm{loc}},\varepsilon_{n}(e_{k+1}),\mathbf{1}_{a_{\mathrm{s},k}=n}) &  \varepsilon_{n}(e_{k+1})=0 \\
	(\tilde{L}_{n,k},\tilde{L}_{n,k}^{\mathrm{loc}}-1,\varepsilon_{n}(e_{k+1}),\mathbf{1}_{a_{\mathrm{s},k}=n}) & \varepsilon_{n}(e_{k+1})=-1 \\
	\end{array}\right..
	\end{align}	
	
	As a remark, note that the cardinality of the local state space $\mathcal{S}_{n}$ for any IoT device $n$ is $|\mathcal{S}_{n}|=(M+1)\times(M^{\mathrm{loc}}+1)\times4\times2$, which does not grow with the number of IoT devices $N$. In contrast, the cardinality of the global state space grows exponentially with the number of IoT devices $N$.\par

	\subsection{Value Function Approximation}	
	First, the local reward function $\tilde{g}_{n}(\tilde{\mathbf{s}}_{k})$ is given as
	\begin{align}
	\label{eq25}
	\tilde{g}_{n}(\tilde{\mathbf{s}}_{k})=&\frac{\omega'_{n}}{\tilde{\beta}(\tilde{\mathbf{s}}_{k})\sum_{l=1}^{N}\lambda_{l}}(\tilde{L}_{n,k}+\tilde{L}_{n,k}^{\mathrm{loc}})+\frac{\gamma'_{n}}{\tilde{\beta}(\tilde{\mathbf{s}}_{k})N}(P_{n}\times \IEEEnonumber \\ &\mathbf{1}_{a_{\mathrm{s},k}=n}+P_{n}^{\mathrm{loc}}\times\mathbf{1}_{\tilde{L}_{n,k}^{\mathrm{loc}}\neq 0}),
	\end{align}
	\noindent so that $\tilde{g}(\tilde{\mathbf{s}}_{k})=\sum_{n=1}^{N}\tilde{g}_{n}(\tilde{\mathbf{s}}_{k})$.\par
	
Moreover, we decompose the optimal average reward rate $\theta$ in \eqref{eq21} as the sum of optimal local average reward rates $\theta_{n}$ of IoT $n\in\mathcal{N}$, i.e.,
	\begin{equation}
    \label{eq53}
    \theta=\sum_{n=1}^{N}\theta_{n}.
	\end{equation}

		\begin{figure}
		\centering
		\includegraphics[width=0.5\textwidth]{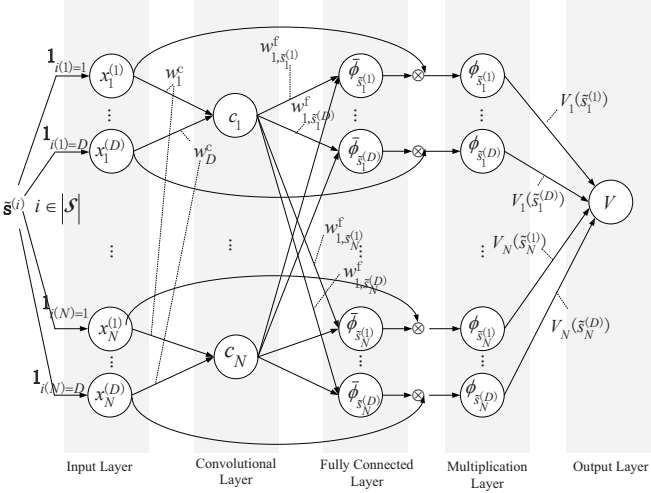}
		\caption{Function Approximation Architecture.}
		\label{vfa}
	\end{figure}
	
	In order to formulate our approximation architecture, we first introduce some notations to efficiently describe the mapping relations between the post-decision global system states and the post-decision local system states. Specifically, denote $\tilde{\mathbf{s}}^{(i)}\in\mathcal{S}$ as the $i$-th post-decision global system state in the state space. We introduce a mapping function $i(n)$ which denotes the index of the post-decision local system state of device $n$ within its local state space when the post-decision global system state is $\tilde{\mathbf{s}}^{(i)}$. Therefore, let $\tilde{s}_{n}^{(i(n))}\in\mathcal{S}_{n}$ denote the local system state of device $n$ when the global system state is $\tilde{\mathbf{s}}^{(i)}$. In other words, we have $\tilde{\mathbf{s}}^{(i)}=\{\tilde{s}_{n}^{(i(n))}\}_{n=1}^{N}$.\par
	
	The approximation architecture for the post-decision global value function is given as
	\begin{align}
	\label{eq26}
	V(\tilde{\mathbf{s}}^{(i)})\cong\sum_{n=1}^{N}\sum_{j=1}^{D}\phi_{\tilde{s}_{n}^{(j)}}(\tilde{\mathbf{s}}^{(i)})V_{n}(\tilde{s}_{n}^{(j)}),
	\end{align}
	\noindent where $D=|\mathcal{S}_{n}|$ is the cardinality of the local system space of any device $n\in\mathcal{N}$, and $V_{n}(\tilde{s}_{n}^{(j)})$ is the post-decision per-node value function of IoT device $n$ for its post-decision local system state $\tilde{s}_{n}^{(j)}$. In the following discussion, we'll omit the term ``post-decision" before the global value function and per-node value function for simplicity.  $\{\phi_{\tilde{s}_{n}^{(j)}}(\tilde{\mathbf{s}}^{(i)})\}_{n=1}^{N}$ is the feature vector of the post-decision global system state $\tilde{\mathbf{s}}^{(i)}$. The simplest method is to set $\phi_{\tilde{s}_{n}^{(j)}}(\tilde{\mathbf{s}}^{(i)})=\mathbf{1}_{j=i(n)}$ for any $n\in\mathcal{N}$. However, the feature values of the local system state $\tilde{s}_{n}^{(j)}$ are the same for all the global system states that $\tilde{s}_{n}^{(j)}$ belongs to, i.e., the component values within the set $\{\phi_{\tilde{s}_{n}^{(j)}}(\tilde{\mathbf{s}}^{(i)})|i\in|\mathcal{S}|,j=i(n)\}$ are the same. This can lead to inaccuracy in the approximation by \eqref{eq26}. In this paper, we will use an ANN to train the values of $\phi_{\tilde{s}_{n}^{(j)}}(\tilde{\mathbf{s}}^{(i)})$ and $V_{n}(\tilde{s}_{n}^{(j)})$ simultaneously as shown in Fig.\ref{vfa}.  \par

	\subsubsection{Input layer}
	The input of the neural network is a post-decision global system state $\tilde{\mathbf{s}}^{(i)}$, $\forall i\in|\mathcal{S}|$. The input layer has $(N\times D)$ neurons, where the $((n-1)D+j)$-th neuron corresponds to the $j$-th post-decision local system state of device $n\in\mathcal{N}$, i.e., $\tilde{s}_{n}^{(j)}$.  
	\begin{itemize}
		\item Activation $x_{n}^{(j)}$: The activation of the neuron $\tilde{s}_{n}^{(j)}$ is denoted by $x_{n}^{(j)}$, which is a $0$-$1$ variable indicating whether $\tilde{s}_{n}^{(j)}$ is a component of the input post-decision global system state $\tilde{\mathbf{s}}^{(i)}$, i.e, $\mathbf{1}_{i(n)=j}$. Let $\mathbf{x}(\tilde{\mathbf{s}}^{(i)})$ be the $(N\times D)$-dimensional matrix of input neurons where the element in the $n$-th row and $j$-th column is $\mathbf{1}_{i(n)=j}$.
	\end{itemize}

	\subsubsection{Convolutional layer}
	The number of neurons in the convolutional layer is $N$. The $n$-th neuron corresponds to the device $n\in \mathcal{N}$.
	\begin{itemize}
	\item Activation $c_{n}$: The activation of the $n$-th neuron is denoted by $c_{n}$, which represents the convolutional layer feature extracted from the local system state $s_{n}^{(i(n))}$ that is encoded by the $n$-th row of $\mathbf{x}(\tilde{\mathbf{s}}^{(i)})$. Let $\mathbf{c}(\tilde{\mathbf{s}}^{(i)})$ be the $N$-dimensional activation vector of the convolutional layer whose $n$-th element is $c_{n}$.
	\item C-weight/Filter $w^{\mathrm{c}}_{j}$: The filter is a $D$-dimensional vector $\mathbf{w^{\mathrm{c}}}=\{w^{\mathrm{c}}_{j}\}_{j=1}^{D}$, where each element $w^{\mathrm{c}}_{j}$ is referred to as the c-weight. 
    \end{itemize}	
		
    Thus, we have
	\begin{equation}
	\label{eq44}
	\mathbf{c}(\tilde{\mathbf{s}}^{(i)})=\mathbf{w^{\mathrm{c}}}\times\mathbf{x}(\tilde{\mathbf{s}}^{(i)})^{\mathrm{T}},
	\end{equation}
    \noindent where $\mathbf{x}(\tilde{\mathbf{s}}^{(i)})^{\mathrm{T}}$ denotes the transpose of the matrix $\mathbf{x}(\tilde{\mathbf{s}}^{(i)})$.\par

	\subsubsection{Fully connected layer}
	The number of neurons in the fully connected layer is the same with that in the input layer, i.e., $N\times D$, where the $((n-1)D+j)$-th neuron corresponds to the post-decision local system state $\tilde{s}_{n}^{(j)}$.
	\begin{itemize}
	\item Activation $\bar{\phi}_{\tilde{s}_{n}^{(j)}}$: The activation of the neuron $\tilde{s}_{n}^{(j)}$ is denoted by $\bar{\phi}_{\tilde{s}_{n}^{(j)}}$.  Let $	\boldsymbol{\bar{\phi}}(\tilde{\mathbf{s}}^{(i)})$ be the $(N\times D)$-dimensional activation vector of the fully connected layer whose $((n-1)D+j)$-th element is $\bar{\phi}_{s_{n}^{(j)}}$. 
	\item F-weight $w_{m,\tilde{s}_{n}^{(j)}}^{\mathrm{f}}$: The f-weight $w_{m,\tilde{s}_{n}^{(j)}}^{\mathrm{f}}$ is the weight of the link from the $m$-th neuron in the convolutional layer to the neuron $\tilde{s}_{n}^{(j)}$ in the fully connected layer. Let $\mathbf{w^{\mathrm{f}}}$ denote the $(N\times(N\times D))$-dimensional matrix of f-weights, whose element in the $m$-th row and the $((n-1)D+j)$-th column is $w_{m,\tilde{s}_{n}^{(j)}}^{\mathrm{f}}$.	
    \end{itemize}
    To derive the activation vector of the fully connected layer, we have	
	\begin{equation}
\label{eq45}
\boldsymbol{\bar{\phi}}(\tilde{\mathbf{s}}^{(i)})=\sigma(\mathbf{c}(\tilde{\mathbf{s}}^{(i)})\times\mathbf{w^{\mathrm{f}}}),
\end{equation}	
\noindent where we set $\sigma$ to be the Tanh function defined by $\sigma(z)=\frac{2}{1+e^{-2z}}-1$. \par
    

	\subsubsection{Multiplication layer}
	Note that when $j\neq i(n)$, we would like $\phi_{\tilde{s}_{n}^{(j)}}(\tilde{\mathbf{s}}^{(i)})=0$ according to \eqref{eq26}. In order to guarantee this, we add a multiplication layer between the fully connected layer and the output layer. There are also $N\times D$ neurons in the multiplication layer.
	\begin{itemize}
	\item Activation $\phi_{\tilde{s}_{n}^{(j)}}$: The activation of the $((n-1)D+j)$-th neuron is the feature $\phi_{\tilde{s}_{n}^{(j)}}$ in \eqref{eq26}. The $(N\times D)$-dimensional activation vector $\boldsymbol{\phi}(\tilde{\mathbf{s}}^{(i)})$ whose $((n-1)D+j)$-th element is $\phi_{\tilde{s}_{n}^{(j)}}$ can be derived by
	\begin{equation}
	\label{eq49}
	\boldsymbol{\phi}(\tilde{\mathbf{s}}^{(i)})=\boldsymbol{\bar{\phi}}(\tilde{\mathbf{s}}^{(i)})\odot\mathbf{x}(\tilde{\mathbf{s}}^{(i)}).
	\end{equation} 	
	\noindent where $\odot$	is the Hadamard product or the elementwise product of two vectors. 

    \end{itemize}	

	\newtheorem{remark}{Remark}
\begin{remark}[Discussion on the multiplication layer]	
 Note that the neural network and the DRL algorithm still work without the multiplication layer. The advantage of the  multiplication layer lies in that it greatly reduces the number of weights or parameters that need to be udpated at each decision epoch. Specifically, among the $N\times D$ neurons in the multiplication layer, only $N$ neurons are active per decision epoch, which means that only the f-weights and per-node value functions associated with these active neurons need to be updated. This is similar to Dropout in NN, which can prevent overfitting of the value function. Therefore, the multiplication layer not only greatly reduces the computation complexity of the DRL algorithm, but also improves its performance. 
\end{remark}

	\subsubsection{Output layer}
	\begin{itemize}
	\item Activation $V(\tilde{\mathbf{s}}^{(i)})$: The output of the neural network is the global value function of the input post-decision global system state $\tilde{\mathbf{s}}^{(i)}$, $\forall i\in|\mathcal{S}|$. Therefore, there is only one output neuron whose activation is $V(\tilde{\mathbf{s}}^{(i)})$. 
	\item Weight/Per-node value function $V_{n}(\tilde{s}_{n}^{(j)})$: The purpose of the output layer is to derive the global value function of $\tilde{\mathbf{s}}^{(i)}$ according to \eqref{eq26}. The per-node value function $V_{n}(\tilde{s}_{n}^{(j)})$ is the weight of the link from the neuron $\tilde{s}_{n}^{(j)}$ in the fully connected layer to the output neuron. Let $\mathbf{V}$ be the $(N\times D)$-dimensional per-node value function (weight) vector with the $((n-1)D+j)$-th element as $V_{n}(\tilde{s}_{n}^{(j)})$.
\end{itemize}	

 Then, \eqref{eq26} can be written as below
\begin{equation}
\label{eq46}
V(\tilde{\mathbf{s}}^{(i)})\cong\mathbf{V}\boldsymbol{\phi}(\tilde{\mathbf{s}}^{(i)}).
\end{equation}			
	
	With \eqref{eq25} and \eqref{eq26}, the Bellman function \eqref{eq21} can be written as \eqref{eq27} on top of the next page.
	\newcounter{mytempeqncnt}
\begin{figure*}
	\normalsize
	\setcounter{mytempeqncnt}{\value{equation}}
	\setcounter{equation}{23}
	\begin{align}
	\label{eq27}
	\sum_{n=1}^{N}\phi_{\tilde{s}_{n,k}}(\tilde{\mathbf{s}}_{k})V_{n}(\tilde{s}_{n,k})=&\sum_{\mathbf{s}_{k+1}\in\mathcal{S}}q(\mathbf{s}_{k+1}|\tilde{\mathbf{s}}_{k})\min_{\pi}\Big(\sum_{n=1}^{N}\tilde{g}_{n}(\tilde{\mathbf{s}}_{k+1})+\sum_{n=1}^{N}\phi_{\tilde{s}_{n,k+1}}(\tilde{\mathbf{s}}_{k+1})V_{n}(\tilde{s}_{n,k+1})-\theta/\tilde{\beta}(\tilde{\mathbf{s}}_{k+1})\Big) ,
	\end{align} 
	\setcounter{equation}{\value{mytempeqncnt}}
	\hrulefill
\end{figure*}

In the following discussion, we will refer to our proposed algorithm as the neural-ICFMO algorithm, which takes the first letter of each neural network layer.

    \subsection{Optimal Control Action}    
    To illustrate the structure of our solution, we first assume that we could obtain the per-node value function vector $\mathbf{V}$, the f-weight matrix $\mathbf{w^{\mathrm{f}}}$, the c-weight vector $\mathbf{w^{\mathrm{c}}}$, and the local optimal average reward rate $\theta_{n}$ for every IoT device $n\in\mathcal{N}$ via somemeans. At the $k$-th decision epoch, we focus on deriving the optimal action $\mathbf{a}_{k}^{*}=\pi^{*}(\mathbf{s}_{k})$ under the current system state $\mathbf{s}_{k}$ to minimize the value of the RHS of \eqref{eq27} as below:
    
    \begin{align}
    \setcounter{equation}{24}	
	\label{eq28}
	&\pi^{*}(\mathbf{s}_{k})=\arg\min_{\pi}\sum_{n=1}^{N} \IEEEnonumber \\
	&\left(\tilde{g}_{n}(\tilde{\mathbf{s}}_{k}) +\phi_{\tilde{s}_{n,k}}(\tilde{\mathbf{s}}_{k})V_{n}(\tilde{s}_{n,k})-\theta_{n}/\tilde{\beta}(\tilde{\mathbf{s}}_{k})\right).
	\end{align}  
	
	Under the system state $\mathbf{s}_{k}$, the action space $\mathcal{A}_{\mathbf{s}_{k}}$ is formed by all the actions $\mathbf{a}_{k}=(a_{\mathrm{o},k},a_{\mathrm{s},k})$ that is the combination of any eligible offloading action $a_{\mathrm{o},k}\in\mathcal{A}_{\mathrm{o},\mathbf{s}_{k}}$ and eligible scheduling action $a_{\mathrm{s},k}\in\mathcal{A}_{\mathrm{s},\mathbf{s}_{k}}$, where $\mathcal{A}_{\mathrm{o},\mathbf{s}_{k}}$ and $\mathcal{A}_{\mathrm{s},\mathbf{s}_{k}}$ are given by \eqref{eq4} and \eqref{eq5}, respectively. The algorithm to determine the optimal control action is given in Algorithm \ref{alg1}. \par 
	
	\begin{algorithm}
		\caption{Determine the optimal action}
		\label{alg1}
		\begin{algorithmic}
			\FOR{each decision epoch $k=0,1,2,\cdots$}
			\STATE{$k \leftarrow k+1$}
			\FOR{each action $\mathbf{a}_{k}\in\mathcal{A}_{\mathbf{s}_{k}}$}
			\STATE Determine the post-decision global system state $\tilde{\mathbf{s}}_{k} \leftarrow f(\mathbf{s}_{k},\mathbf{a}_{k})$ by \eqref{eq8}
			\STATE Determine the post-decision local system state vector $\{\tilde{s}_{n,k}\}_{n=1}^{N} \leftarrow \{f_{n}(s_{n,k},\mathbf{a}_{k})\}_{n=1}^{N}$ by \eqref{eq24}
			\FOR{each IoT device $n\in\mathcal{N}$}
			\STATE Encode $\tilde{\mathbf{s}}_{k}$ in the input vector $\mathbf{x}_{\tilde{s}_{n,k}}$
			\STATE Determine the feature $\phi_{\tilde{s}_{n,k}}(\tilde{\mathbf{s}}_{k})$ by \eqref{eq44}, \eqref{eq45}, \eqref{eq49}
			\STATE Determine the local reward function $\tilde{g}_{n}(\tilde{\mathbf{s}}_{k})$ by \eqref{eq25} 
			\ENDFOR
			\STATE 	Determine the value of the RHS of \eqref{eq28}
			\ENDFOR
			\STATE Select the action according to \eqref{eq28}
			\ENDFOR
		\end{algorithmic}
	\end{algorithm}

    \subsection{Per-Node Value Function and Weight Update}	
     In the above discussion, we consider that all the per-node value functions, f-weights, c-weights, and $\theta_{n}$ are known for every IoT device, and derive the optimal control actions based on these values. In this section, we will discuss how to derive the above values using the stochastic gradient (SGD) TD(0) method under function approximation for the average-reward problem \cite{Sutton:book}. The loss function at the $k$-th decision epoch is defined as
     \begin{align}
     \label{eq48}
     &L_{k}(\mathbf{V}_{k},\mathbf{w^{\mathrm{f}}}_{k},\mathbf{w^{\mathrm{c}}}_{k})=\frac{1}{2}\mathbb{E}\Bigg(\min_{\pi}\sum_{n=1}^{N}\Big(\tilde{g}_{n}(\tilde{\mathbf{s}}_{k})+\phi_{\tilde{s}_{n,k},k}(\tilde{\mathbf{s}}_{k}) \IEEEnonumber \\ &V_{n,k}(\tilde{s}_{n,k})-\phi_{\tilde{s}_{n,k-1},k}(\tilde{\mathbf{s}}_{k-1})V_{n,k}(\tilde{s}_{n,k-1})-\theta_{n,k}/\tilde{\beta}(\tilde{\mathbf{s}}_{k})\Big)\Bigg)^{2},
     \end{align}
     \noindent where $\theta_{n,k}$ is the average reward rate of IoT device $n$ up to the $(k-1)$-th decision epoch. The gradient for the loss function is derived by the well-known backpropagation algorithm for the ANN. The detailed procedures and equations to update the value functions and weights are given in Appendix B and summarized in Algorithm \ref{alg2} below. Note that at the $k$-th decision epoch, $\theta_{n,k}$ is used in place of $\theta_{n}$  in \eqref{eq28} to derive the optimal action $\pi^{*}(\mathbf{s}_{k})$. In the following discussion, we add a subscript $k$ to the notations described in Section III.B to represent the parameter values at the $k$-th decision epoch. \par

    \begin{algorithm}
    	\caption{Deep reinforcement learning algorithm}
    	\label{alg2}
    	\begin{algorithmic}
    		\STATE Initialize $\mathbf{V}_{0}$, $\mathbf{w^{\mathrm{f}}}_{0}$, $\mathbf{w^{\mathrm{c}}}_{0}$, and $\{\theta_{n,0}\}_{n=1}^{N}$.
    		\FOR{each decision epoch $k=0,1,2,\cdots$}
    		\STATE{$k \leftarrow k+1$}
    		\STATE Take action $\mathbf{a}_{k}$ determined by Algorithm 1.
    		\STATE $\tilde{\mathbf{s}}_{k} \leftarrow f(\tilde{\mathbf{s}}_{k},\mathbf{a}_{k})$ by \eqref{eq8}
    		\FOR{each IoT device $n=1,2,\cdots,N$}
    		\STATE $\tilde{s}_{n,k} \leftarrow f_{n}(s_{n,k},\mathbf{a}_{k})$ by \eqref{eq24}
    		\ENDFOR
    		\FOR{each IoT device $n=1,2,\cdots,N$}
    		\STATE Update the per-node value function $V_{n,k}(\tilde{s}_{n,k-1})$ by \eqref{eq34},\eqref{eq35},\eqref{eq36}
    		\STATE Update the f-weight vector $\mathbf{w^{\mathrm{f}}}_{\tilde{s}_{n,k-1},k}$ by \eqref{eq47},\eqref{eq50}
    		\STATE Update the c-weight vector $\mathbf{w^{\mathrm{c}}}_{\tilde{s}_{n,k-1},k}$ by \eqref{eq51},\eqref{eq52}    		
    		\STATE Update the average reward $\theta_{n,k}$ by \eqref{eq33},\eqref{eq42},\eqref{eq43}
    		\ENDFOR
    		\STATE Generate the next event $e_{k+1}$
    		\FOR{each IoT device $n=1,2,\cdots,N$}
    		\STATE $s_{n,k+1} \leftarrow h_{n}(\tilde{s}_{n,k},e_{k+1})$ according to \eqref{eq41}
    		\ENDFOR
    		\ENDFOR
    	\end{algorithmic}
    \end{algorithm}

    \subsection{Semi-Distributed Implementation of the Solution}
\begin{figure}[!htb]
    	\centering
    	\includegraphics[width=0.5\textwidth]{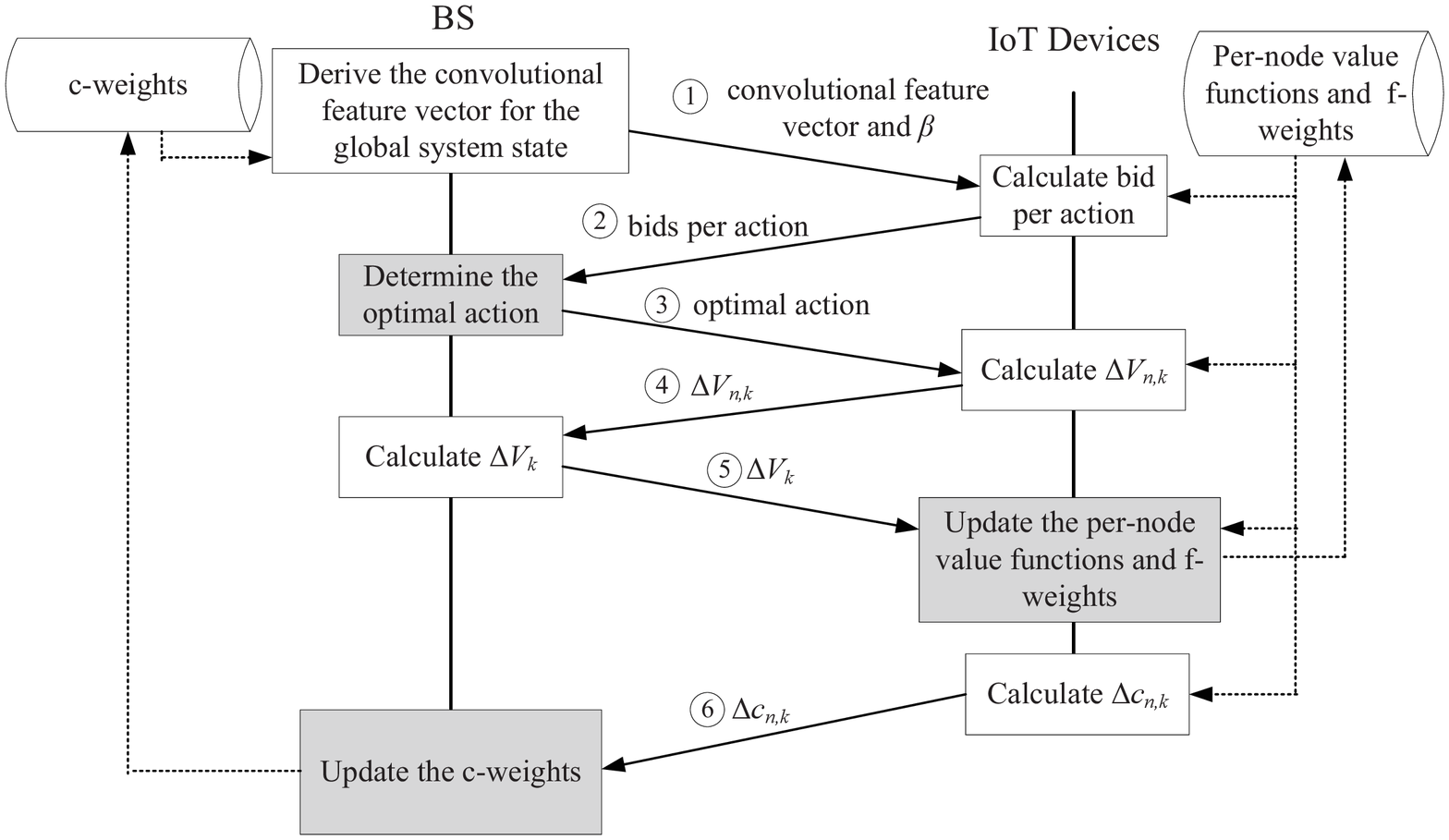}
    	\caption{Signaling of the semi-distributed implementation.}
    	\label{signaling}
    \end{figure}

    The proposed deep reinforcement learning algorithm (i.e., Algorithm 2) can be implemented centrally at the BS. In this case, the BS needs to store the per-node value function vectors, the f-weights and the c-weights for all the IoT devices, whose number grows quadratically instead of exponentially with the number of IoT devices $N$ due to the function approximation. Moreover, all the computational tasks for deriving control actions and maintaining the per-node value function vectors, the f-weights and c-weights need to be performed at the BS. On the other hand, the proposed algorithm also allows semi-distributed implementation, in which the BS and IoT devices collaboratively determine the optimal policy as illustrated in Fig.\ref{signaling}. \par
    
    Specifically, we consider that each IoT device $n\in\mathcal{N}$ stores and updates its own per-node value function vector $\{V_{n,k}(\tilde{s}_{n}^{(j)})\}_{j=1}^{D}$ and the f-weight vectors  $\mathbf{w^{\mathrm{f}}}_{\tilde{s}_{n}^{(j)},k}=\{w_{m,\tilde{s}_{n}^{(j)}}^{\mathrm{f}}\}_{m=1}^{N}$, $\forall j\in\{1,\cdots,D\}$. Note that the $N$-dimensional local f-weight vector $\mathbf{w^{\mathrm{f}}}_{\tilde{s}_{n}^{(j)},k}$ of IoT device $n$ consists of the f-weights from any convolutional neuron $c_{m}$, $m\in\mathcal{N}$ to its local system state $\tilde{s}_{n}^{(j)}$. On the other hand, the BS stores and updates the c-weight vector $\mathbf{w^{\mathrm{c}}}$. At the beginning of the $k$-th decision epoch, the BS first derives the convolutional feature vector $\mathbf{c}_{k}$ of the global system state $\mathbf{s}_{k}$, which corresponds to the activation vector of the convolutional layer of the proposed NN. The BS broadcasts the vector and the value of $\tilde{\beta}(\tilde{\mathbf{s}}_{k})$ to the IoT devices, where every IoT device derives a bid $\tilde{g}_{n}(\tilde{\mathbf{s}}_{k})+\phi_{\tilde{s}_{n,k}}(\tilde{\mathbf{s}}_{k})V_{n}(\tilde{s}_{n,k})$ for each eligible action under the current system state. The IoT devices submit the bids to the BS, which determine the optimal control action by \eqref{eq28} based on the bid values. The BS notifies the optimal action to the IoT devices, which calculate the values of $\Delta V_{n,k}(\tilde{s}_{n,k-1})$ by \eqref{eq36} and submit them to the BS. Then, the BS derives the value of $\Delta V_{k}(\tilde{\mathbf{s}}_{k-1})$ by \eqref{eq35} and broadcasts the value to the IoT devices, which updates their per-node value functions and f-weights, respectively. Finally, every IoT device $n$ derives the vector $\Delta \mathbf{c}_{n,k}$ by \eqref{eq52} and submits the information to the BS, while the BS updates the c-weights according to \eqref{eq51}.\par

	\newtheorem{remark2}[remark]{Remark}
\begin{remark2}[Signaling overhead of semi-distributed implementation]	
The signaling overhead associated with each message in Fig. \ref{signaling} is listed in Table \ref{overhead}. The signaling overhead grows linearly with the number of IoT devices $N$. Note that message \textcircled{3} is needed irrespective of the implementation method. Messages \textcircled{4}, \textcircled{5}, and \textcircled{6} are only needed in the learning phase. After the convergence of the parameters in the neural network, only messages \textcircled{1} and \textcircled{2} are needed. The frequency of the signaling exchange depends on the packet arrival and departure duration of NB-IoT applications, which is typically much longer than the $1$ ms time slot in LTE system.    
\end{remark2}

    \begin{table}[!htb]
	\renewcommand{\arraystretch}{1.3}
	\caption{Signaling overhead} \label{overhead} \centering
	\begin{tabular}{|c|c|c|c|c|c|c|}
		\hline
		\textbf{Message} & \textcircled{1}  & \textcircled{2} & \textcircled{3} & \textcircled{4} & \textcircled{5} & \textcircled{6} \\
		\hline
		\textbf{Overhead ($\times4$bytes)} & $(N+1)$  & $2N$ & $1$ & $N$ & $1$ & $N$  \\
		\hline
	\end{tabular}
\end{table}  
    
    \section{Simulation Results}
    In this section, we evaluate the performance of our proposed computation offloading and user scheduling algorithm using DRL, i.e., neural-ICFMO, via simulation. Specifically, we compare its performance with 
    the following algorithms:
    \begin{itemize}
    	\item Queue-aware (QA) algorithm: Both the offloading and scheduling algorithms take into account the queue length and consider load balancing. Specifically, if the transmission queue is shorter than the processing queue at an IoT device, the arrived packet will be offloaded, and vice versa. On the other hand, the IoT device with the largest transmission queue length will be scheduled for transmission. 
    	\item MUMTO algorithm: We adapt the MUMTO algorithm in \cite{ICC:Chen}, which is a typical offloading and resource allocation algorithm designed based on the deterministic task model. Specifically, the optimization objective at the $k$-th decision epoch is $\min(\sum_{n=1}^{N}\omega_{n}D_{n}(a_{k})+\sum_{n=1}^{N}\gamma_{n}P_{n}(a_{k}))$, where $D_{n}(a_{k})$ and $P_{n}(a_{k})$ are the packet transmission/processing delay and power consumption of IoT device $n$ if action $a_{k}$ is selected at the $k$-th decision epoch. 
    	\item Neural-ICO algorithm: It is similar to the DQN algorithm used in \cite{Chen:IoT,He:TET}, except that the output from the neural network is the value functions for the post-decision states instead of the Q factors for the state and action pairs. The function approximation architecture of this algorithm is given in Fig.\ref{ico} in Appendix C.
    \end{itemize}

    We developed a discrete-event system-level simulator for the NB-IoT MEC system, where the the simulation parameters are given in Table \ref{para}. We simulate a circular cell of radius $R$ with a BS in the center. The IoT devices are uniformly distributed in the cell area in random. We divide the circular area of the considered cell into $K$ disjunct zones by $K-1$ concentric circles around the BS, where the zone $k\in\{1,\ldots,K\}$ is the region between two concentric circles with radius $d_{k-1}$ and $d_{k}$. We consider that all the IoT devices in zone $k$, $k\in\{1,\ldots,K\}$ select the same MCS level from the $13$ MCS levels of the NB-IoT uplink data channel, so that their transmission data rates $r_{k}$ can be determined according to the Table V in \cite{VTC:Ratasuk}. Then, the mean transmission rate in terms of packets per second for IoT device $n$ can be derived as $\mu_{k}=r_{k}/l_{\mathrm{p}}$, where $l_{\mathrm{p}}$ is the mean packet size. Note that the MCS level selection schemes are not the focus of this paper, and our proposed offloading and scheduling algorithm can be used for any link adaptation schemes. The transmission power of the IoT device $n$ is derived according to the NB-IoT uplink open loop power control formulas given in \cite{Access:Yu2}. The CPU cycles $f_{n}^{\mathrm{loc}}$ of the IoT devices are considered to be uniformly distributed. Therefore, the mean local processing rate $\mu_{n}^{\mathrm{loc}}$ and local power consumption $P_{n}^{\mathrm{loc}}$ can be derived according to (1) in \cite{Survey:Mao} and (4) in \cite{TWC:Mao}, respectively. Without loss of generality, we set the weights of delay and power consumption to be $\omega'_{n}=\gamma'_{n}=0.5$ in the reward function. For both the proposed Neural-ICFMO algorithm and the Neural-ICO algorithm, we use the $\varepsilon$-greedy algorithm to explore the actions. Specifically, the optimal action derived by Algorithm 1 is selected with probability $(1-\mathrm{p}(k))$ at the $k$-th decision epoch, where $\mathrm{p}(k)=\frac{G_{1}}{G_{2}+k}$ with $G_{1}=1000$ and $G_{2}=2000$. \par 
    

    \begin{table}[!htb]
    	\renewcommand{\arraystretch}{1.3}
    	\caption{Simulation Parameters} \label{para} \centering
    	\begin{tabular}{|>{\centering}m{4.5cm} |c|}
    		\hline
    		\textbf{Parameter} & \textbf{Value} \\
    		\hline
     	    Cell radius $R$ & $500$ m \\
    		\hline
    		Zone number $K$ & $10$  \\
    		\hline
    		Path-loss channel model & $15.3+37.6 \log_{10}l_{k}[m]$ \\
	
    	     & $l_{k}=\frac{d_{k-1}+d_{k}}{2}$ m  \\
    	    \hline
    	    Maximum transmission power $P_{\mathrm{CMAX}}$ & $23$ dBm \\
    		\hline
    		Number of CPU cycles for processing one bit of data $X$  & $10^{5}$ \\
    		\hline
    		CPU at the IoT device $n$ $f_{n}^{\mathrm{loc}}$ & $[1,3]\times10^{9}$ cycles/sec \\
    		\hline
    		The effective switched capacitance of the
    		CPU $\kappa$  & $10^{-28}$ \\
    		\hline
    		Mean packet size $l_{\mathrm{p}}$ & $10$ kbits \\
    		\hline
    		Maximum queue length $M$, $M^{\mathrm{loc}}$ & 7 packets \\
    		\hline
    	\end{tabular}
    \end{table}  	
    

%
%
%
%
%

\subsection{Performance vs. varying IoT device numbers}
\begin{figure}[!htb]
	\centering
	\includegraphics[width=0.45\textwidth]{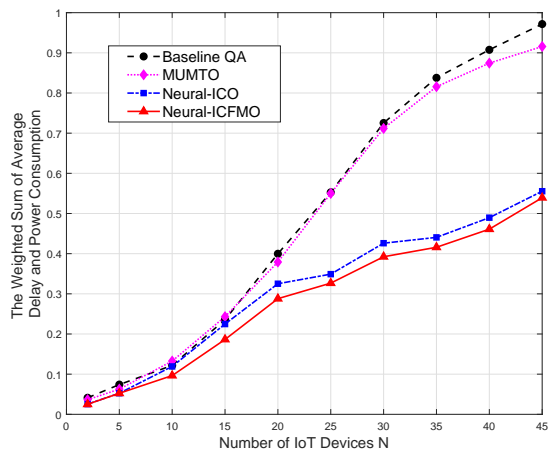}
	\caption{Weighted sum of the average delay and power consumption versus the number of IoT devices $N$ with $\lambda=1$pkts/s.}
	\label{number_weight}
\end{figure}

We set the packet arrival rate $\lambda=1$ pkts/s. The baseline algorithm QA, the MUMTO algorithm, the proposed Neural-ICFMO algorithm, and the Neural-ICO algorithm are simulated when the number of IoT devices $N$ is varied from $2$ to $45$. From Fig.\ref{number_weight}, it can be observed that the proposed Neural-ICFMO algorithm constantly outperforms the other three algorithms in terms of the weighted sum of delay and power consumption, which is the optimization objective as given in \eqref{eq12} for both the Neural-ICFMO and Neural-ICO algorithms. The performance of the QA and the MUMTO algorithms are quite similar, and the performance improvements of the proposed Neural-ICFMO algorithm as well as the Neural-ICO algorithm over these two algorithms increase with the increasing IoT device number. This is because the MUMTO algorithm solves a static optimization problem independently at each decision epoch to optimize the current performance, without considering the impact of the current decision to the long-term time-average performance. Therefore, its performance is not as good as those of the Neural-ICFMO and Neural-ICO algorithms. The proposed Neural-ICFMO algorithm also outperforms the Neural-ICO algorithm due to its neural network architecture for function approximation. Specifically, the neurons in the fully connected layer of the Neural-ICFMO algorithm correspond to the local system states of the IoT devices. This design principle not only facilitates its semi-distributed implementation, but also improves its performance compared with the Neural-ICO algorithm. \par

\begin{figure}[!htb]
	\centering
	\includegraphics[width=0.45\textwidth]{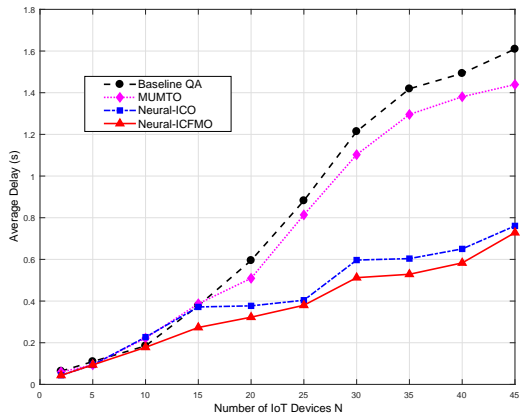}
	\caption{Average delay versus the number of IoT devices $N$ with $\lambda=1$pkts/s.}
	\label{number_delay}
\end{figure}

\begin{figure}[!htb]
	\centering
	\includegraphics[width=0.45\textwidth]{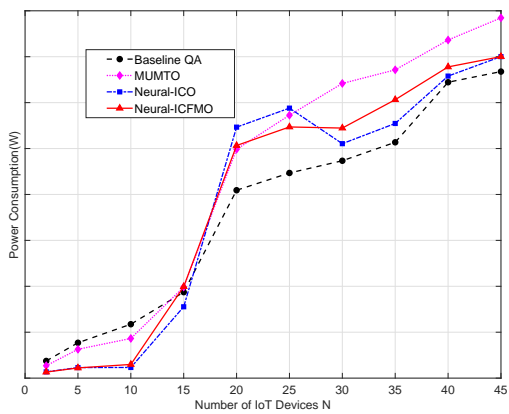}
	\caption{Power consumption versus the number of IoT devices $N$ with $\lambda=1$pkts/s.}
	\label{number_power}
\end{figure}

Fig.\ref{number_delay} and Fig.\ref{number_power} compare the average delay and the power consumption of the four algorithms with increasing IoT device number, respectively. It can be observed that the proposed Neural-ICFMO algorithm achieves the best delay performance, but not the best power consumption performance. This is because the delay value is larger than the power consumption value in our configuration, so that the Neural-ICFMO sacrifices a small amount of power consumption performance to improve a larger amount of delay performance, so that it can achieve the best tradeoff in minimizing the weighted sum of the delay and power consumption. It can also be observed from Fig.\ref{number_power} that the power consumption values can decrease with the increasing IoT device number for the Neural-ICFMO and the Neural-ICO algorithms. For example, the power consumption of Neural-ICFMO is $0.2735$ W when $N=25$ and $0.2723$ W when $N=30$. This is because in order to optimize the weighted sum of delay and power consumption, both the Neural-ICFMO and Neural-ICO algorithms may choose to give more priority to minimizing delay or power consumption depending on the specific configuration. Note that the weighted sum of delay and power consumption increases with the increasing IoT device number as given in Fig.\ref{number_weight}, which makes sense as the increasing IoT device number leads to increasing traffic load and decreasing performance.\par

\subsection{Performance vs. varying packet arrival rates}
\begin{figure}[!htb]
	\centering
	\includegraphics[width=0.45\textwidth]{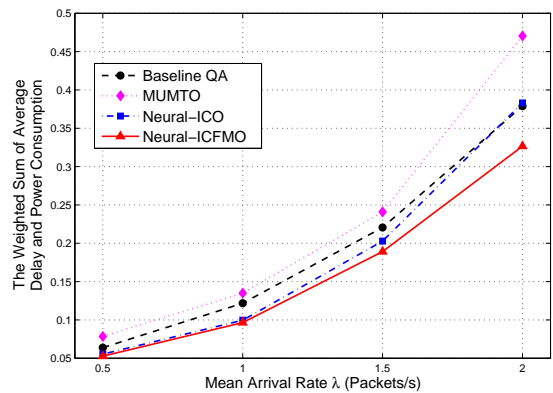}
	\caption{Weighted sum of the average delay and power consumption versus the mean arrival rate $\lambda$ with $\omega'_{n}=\gamma'_{n}=0.5$ and $N=20$.}
	\label{arrival_weight}
\end{figure}

We set the number of IoT devices to be $N=10$. The four algorithms discussed above are simulated when the packet arrival rate $\lambda$ is varied from $0.5$ pkts/s to $2$ pkts/s. From Fig.\ref{arrival_weight}, it can be observed that the proposed Neural-ICFMO algorithm performs constantly better than the other three algorithms with various packet arrival rates in terms of the weighted sum of delay and power consumption. The baseline QA algorithm outperforms the MUMTO algorithm in this configuration, which demonstrates that load balance is a good strategy. The performance of the Neural-ICO algorithm is approximately the same with that of the proposed Neural-ICFMO algorithm when the arrival rates are small. However, the performance gap between the two algorithms increases with the increasing arrival rate. \par

\begin{figure}[!htb]
	\centering
	\includegraphics[width=0.45\textwidth]{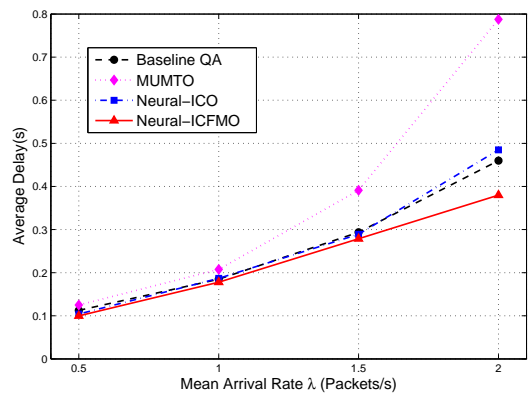}
	\caption{Average delay versus the mean arrival rate $\lambda$ with $N=10$.}
	\label{arrival_delay}
\end{figure}

\begin{figure}[!htb]
	\centering
	\includegraphics[width=0.45\textwidth]{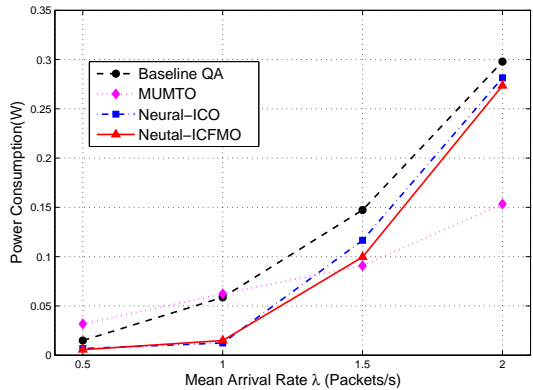}
	\caption{Power consumption versus the mean arrival rate $\lambda$ with $N=10$.}
	\label{arrival_power}
\end{figure}

Fig.\ref{arrival_delay} and Fig.\ref{arrival_power} compare the average delay and the power consumption of the four algorithms with increasing arrival rate, respectively. It can be observed that the proposed Neural-ICFMO algorithm achieves the best delay performance and the best power performance under small packet arrival rates. When the packet arrival rate is larger than $1$ pkts/s, the MUMTO algorithm achieves the best power performance. However, the delay performance of the MUMTO algorithm is the worst among all the algorithms, and its weighted sum of delay and power consumption is the largest. \par

\subsection{Convergence}
\begin{figure}[!htb]
	\centering
	\includegraphics[width=0.45\textwidth]{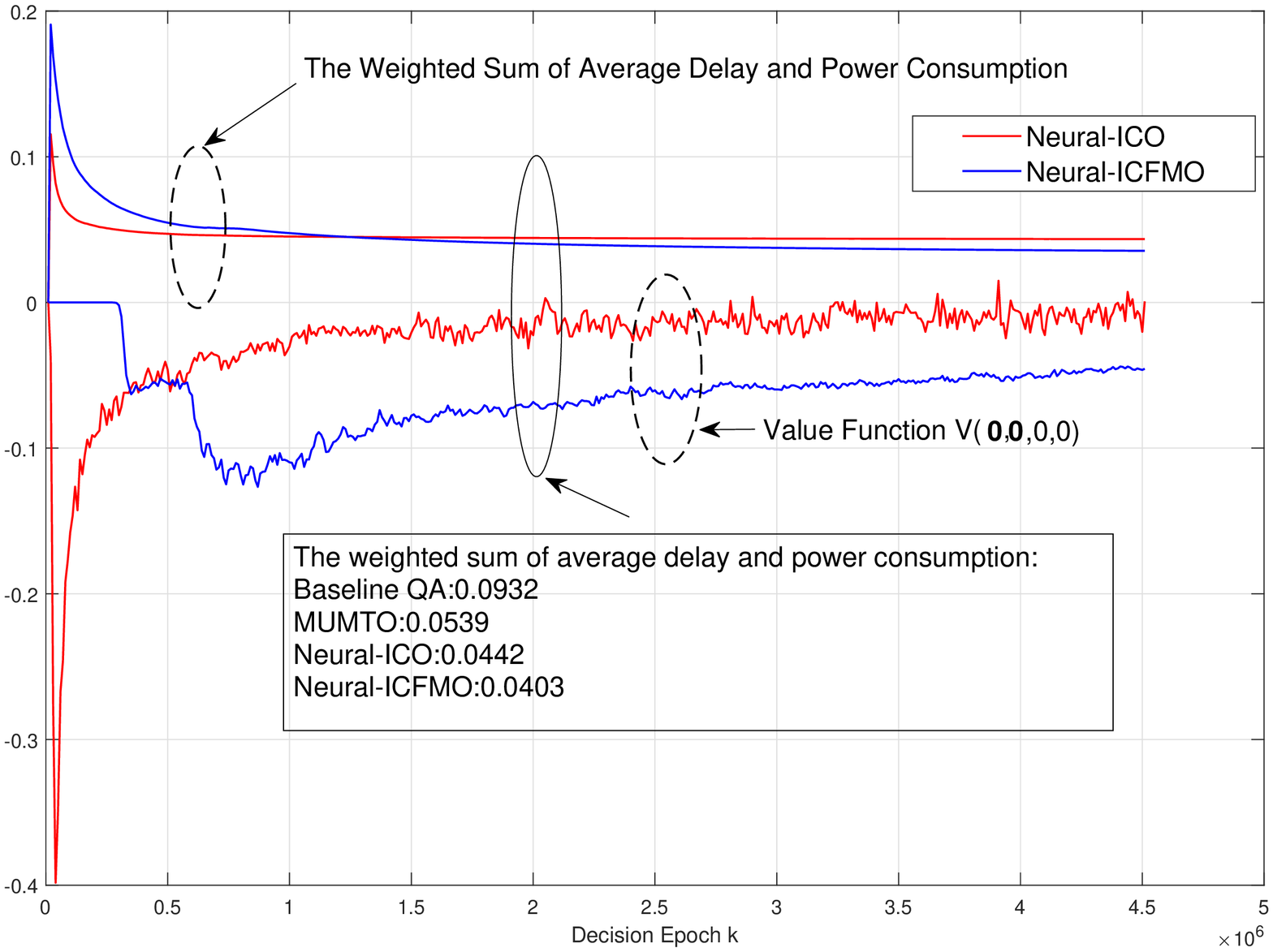}
	\caption{Convergence property of the Neural-ICFMO algorithm with $N=2$ and $\lambda=1$ pkts/s}
	\label{convergence}
\end{figure}

Fig.\ref{convergence} shows the convergence property of the proposed Neural-ICFMO algorithm and the Neural-ICO algorithm. We plot the value function of global state $(\mathbf{0},\mathbf{0},0,0)$ and the weighted sum of delay and power consumption performance of $2$ IoT devices versus the number of decision epochs at a mean arrive rate $\lambda=1$ pkts/s. It can be seen that the value functions of Neural-ICFMO and Neural-ICO algorithms converge at approximately the same speed while their performance improve with increasing number of decision epochs. The performance of Neural-ICFMO is slightly better than that of the Neural-ICO after convergence. The weighted sum of average delay and power consumption for both Neural-ICFMO and Neural-ICO algorithms at the $2\mathrm{e}6$th decision epoch are already smaller than those of the baseline QA and MUMTO algorithms. \par

%

\section{Conclusion}
In this paper, we have proposed a semi-distributed joint computation offloading and multi-user scheduling algorithm based on DRL in NB-IoT edge computing system to minimize the weighted sum of average delay and power consumption over all the IoT devices. Specifically, we consider the stochastic arrival model and formulate the optimization problem into an infinite horizon average reward CTMDP problem. The CTMDP model is based on post-decision states to obtain model-free solution without requiring the knowledge of the underlying stochastic process. In order to deal with the curse-of-dimensionality problem, we use DRL techniques and propose a neural network architecture for approximating the value functions of the post-decision states. The function approximation architecture enables auction-based semi-distributed implementation of the proposed algorithm. Simulation results demonstrate that the proposed algorithm not only outperforms the baseline algorithm, but also RL algorithms based on other function approximation architectures. \par

\appendix

\subsection{Derivation of reward rate}
According to Little's Law, the average delay of IoT device $n$ can be derived as
\begin{align}
\label{eq15}
\bar{D}_{n}&=p_{n,\mathrm{off}}\bar{D}_{n}^{\mathrm{rem}}+p_{n,\mathrm{noff}}\bar{D}_{n}^{\mathrm{loc}} \\ \nonumber
&=\mathbf{E}^{\pi(\Omega)}[\frac{\tilde{Q}_{n}+\tilde{Q}_{n}^{\mathrm{loc}}}{\lambda_{n}}],
\end{align}
\noindent where $p_{n,\mathrm{noff}}$ and $p_{n,\mathrm{off}}$ denote the local processing (no offloading) probability and the offloading probability of IoT device $n$, respectively.\par

The average power consumption of IoT device $n$ can be derived as the sum of the average local processing power consumption and the average transmission power consumption for remote processing,
\begin{equation}
\label{eq16}
\bar{P}_{n} =P_{n}\times\mathrm{Pr.}(a_{\mathrm{s},k}=n)+P_{n}^{\mathrm{loc}}\times\mathrm{Pr.}(\tilde{Q}_{n,k}^{\mathrm{loc}}\neq 0),
\end{equation}
\noindent where $\mathrm{Pr.}(a_{\mathrm{s},k}=n)$ represents the probability of IoT device $n$ being scheduled so that it will transmit with a power of $P_{n}$. On the other hand, $\mathrm{Pr.}(\tilde{Q}_{n,k}^{\mathrm{loc}}\neq 0)$ represents the probability that the post-decision processing queue states of IoT device $n$ is not zero so that it will process the packets with a power of $P_{n}^{\mathrm{loc}}$.\par

By combining \eqref{eq15} and \eqref{eq16} with \eqref{eq12}, we can derive the expression of $c(\mathbf{s}_{k},\Omega(\mathbf{s}_{k}))$ as given in \eqref{eq19}.\par

\subsection{Procedure for per-node value function and weight updates}
\paragraph{Step 1: Per-node value function update} At the $k$-th decision epoch, after the optimal control action is determined, the per-node value function vector $\{V_{n,k}(\tilde{s}_{n,k-1})\}_{n=1}^{N}$ is updated based on the observation of the post-decision global and local system states at the $(k-1)$-th and $k$-th decision epoch as given below:
\begin{align}
\label{eq34}
&V_{n,k+1}(\tilde{s}_{n,k-1})=V_{n,k}(\tilde{s}_{n,k-1}) \IEEEnonumber \\
&+\epsilon_{\nu(k,n)}\phi_{\tilde{s}_{n,k-1},k}(\tilde{\mathbf{s}}_{k-1})\Delta V_{k}(\tilde{\mathbf{s}}_{k-1}),
\end{align}
\noindent where
\begin{equation}
\label{eq35}
\Delta V_{k}(\tilde{\mathbf{s}}_{k-1})=\sum_{n=1}^{N}\Delta V_{n,k}(\tilde{\mathbf{s}}_{k-1}),
\end{equation}
\noindent and
\begin{align}
\label{eq36}
&\Delta V_{n,k}(\tilde{\mathbf{s}}_{k-1})=\tilde{g}_{n}(\tilde{\mathbf{s}}_{k})+\eta \phi_{\tilde{s}_{n,k},k}(\tilde{\mathbf{s}}_{k})V_{n,k}(\tilde{s}_{n,k}) \IEEEnonumber \\
&-\phi_{\tilde{s}_{n,k-1},k}(\tilde{\mathbf{s}}_{k-1})V_{n,k}(\tilde{s}_{n,k-1})-\theta_{n,k}/\tilde{\beta}(\tilde{\mathbf{s}}_{k}),
\end{align}
\noindent where $0<\eta<1$ is a value as close to 1 as possible, e.g., $\eta=0.99$, and enables the algorithm to converge gracefully to the optimal solution. $\nu(k,n)$ is the number of visits to post-decision local system state $\tilde{s}_{n,k-1}$ for IoT device $n$ up to the $(k-1)$-th decision epoch, i.e., $\nu(k,n)=\Sigma_{k'=0}^{k-1}\mathbf{1}_{\tilde{s}_{n,k'}=\tilde{s}_{n,k-1}}$.\par


\paragraph{Step 2: F-weight vector update}   
For every device $n\in\mathcal{N}$, update the $N$-dimensional f-weight vector $\mathbf{w^{\mathrm{f}}}_{\tilde{s}_{n,k-1},k}$ for the links from all the convolutional neurons to the fully connected layer neuron $\tilde{s}_{n,k-1}$, which corresponds to the transpose of the $((n-1)D+j)$-th column vector of the matrix $\mathbf{w^{\mathrm{f}}}$. 
\begin{equation}
\label{eq47}
\mathbf{w^{\mathrm{f}}}_{\tilde{s}_{n,k-1},k+1}=\mathbf{w^{\mathrm{f}}}_{\tilde{s}_{n,k-1},k}+\epsilon_{\nu(k,n)}\Delta w_{\tilde{s}_{n,k-1},k}\mathbf{c}_{k}(\tilde{s}_{n,k-1}),
\end{equation}
\noindent where 
\begin{align}
\label{eq50}
&\Delta w_{\tilde{s}_{n,k-1},k}= \IEEEnonumber \\
&\Delta V_{k}(\tilde{\mathbf{s}}_{k-1})V_{n,k}(\tilde{s}_{n,k-1})\sigma'(\mathbf{c}_{k}(\tilde{\mathbf{s}}_{k-1})\times\mathbf{w^{\mathrm{f}}}_{\tilde{s}_{n,k-1},k}) 
\end{align}
\noindent and $\sigma'(z)=\frac{4e^{-2z}}{(1+e^{-2z})^{2}}$ is the derivative of the Tanh function. \par

\paragraph{Step 3: C-weight vector update}  
The c-weight vector is updated as below: 
\begin{equation}
\label{eq51}
\mathbf{w^{\mathrm{c}}}_{k+1}=\mathbf{w^{\mathrm{c}}}_{k}+\epsilon\sum_{n=1}^{N} \Delta\mathbf{c}_{n,k}\mathbf{x}(\tilde{\mathbf{s}}_{k-1}),
\end{equation}
\noindent where $\Delta\mathbf{c}_{n}$ is an $N$-dimensional vector derived by
\begin{equation}
\label{eq52}  
\Delta\mathbf{c}_{n,k}=\Delta w_{\tilde{s}_{n,k-1},k}\mathbf{w^{\mathrm{f}}}_{\tilde{s}_{n,k-1},k}
\end{equation}
\paragraph{Step 4: Average reward update} 
The average cost rate at IoT device $n$ up to the beginning of the $k$-th decision epoch, i.e., $\theta_{n,k+1}$, is updated as below:
\begin{equation}
\label{eq33}
\theta_{n,k+1}=(1-\alpha_{k})\theta_{n,k}+\alpha_{k}(\frac{\tilde{g}_{n,k}^{\mathrm{tot}}}{t_{k}^{\mathrm{tot}}}),
\end{equation}
\noindent 
\begin{equation}  
\label{eq42}  
\tilde{g}_{n,k}^{\mathrm{tot}}= \tilde{g}_{n,k-1}^{\mathrm{tot}}+\tilde{g}_{n}(\tilde{\mathbf{s}}_{k-1}),
\end{equation}  
\begin{equation} 
\label{eq43}    
t_{k}^{\mathrm{tot}}=t_{k-1}^{\mathrm{tot}}+t_{k-1}.
\end{equation}              
\noindent where  $t_{k-1}$ is the duration of the $(k-1)$-th decision epoch. $\tilde{g}_{n,k}^{\mathrm{tot}}$ and $t_{k}^{\mathrm{tot}}$ are the total accumulative reward and total time up to the beginning of the $k$-th decision epoch, respectively.

Note that $\{\epsilon_{k}\}$ and $\{\alpha_{k}\}$ are sequences of step sizes that both satisfy
\begin{displaymath}
\lim_{k\rightarrow\infty}x_{k}=0,x_{k}>0,\sum_{k=0}^{\infty}x_{k}=\infty,\sum_{k=0}^{\infty}x_{k}^{2}=0,
\end{displaymath}
\noindent where $x_{k}$ represents either $\epsilon_{k}$ or $\alpha_{k}$. Moreover, we must make sure that $\alpha_{k}$ converges to $0$ faster than $\epsilon_{k}$. One example that satisfies this condition is $\epsilon_{k}=\log(k)/k$ and $\alpha_{k}=9000/(10000+k)$. Ideally, we have $\lim_{k\rightarrow\infty}\alpha_{k}/\epsilon_{k}=0$.\par

\subsection{Function Approximation Architecture for the Neural-ICO algorithm}
\begin{figure}
	\centering
	\includegraphics[width=0.4\textwidth]{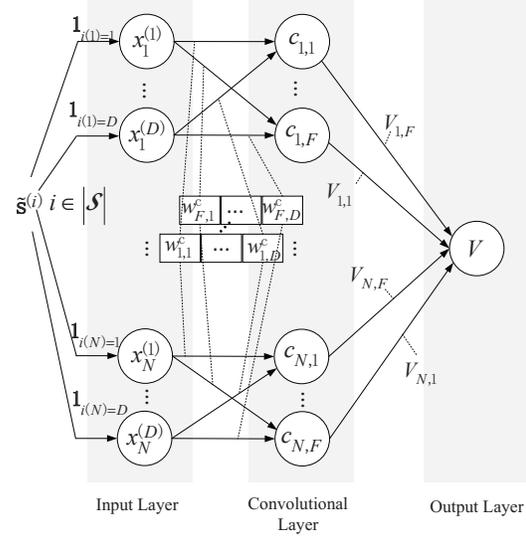}
	\caption{Function Approximation Architecture for neural-ICO algorithm.}
	\label{ico}
\end{figure}

The function approximation architecture for the Neural-ICO algorithm is given in Fig.\ref{ico}. Note that we have one convolutional layer between the input layer and output layer. We have tried to add a fully connected layer after the convolutional layer, but it turns out that this additional layer does not contribute to improving the performance. \par

\end{document}